Time Reversal Invariance in Quantum Mechanics

by

Reza Moulavi Ardakani, M.A, B.S

A Thesis

In

Physics

Submitted to the Graduate Faculty
of Texas Tech University in
Partial Fulfillment of
the Requirements for
the Degree of

MASTER OF SCIENCE

Approved

Dr. Nural Akchurin
Chair of Committee

Dr. Joel Velasco

Dr. Mahdi Sanati

Mark Sheridan
Dean of the Graduate School

December, 2017

# ACKNOWLEDGMENTS

I would like to thank my supervisors, Nural Akchurin, Mahdi Sanati, and Joel Velasco for guiding me while writing this thesis and also for the courses they taught me over the last three years. I am particularly grateful for the assistance they provided by listening to my ideas and questions, and even more for understanding my special mental health situation and their patience with me throughout my study.

I would especially like to thank Nural for engaging me with new ideas and providing me with the opportunity to choose this amazing topic. I offer my most heartfelt thanks to his kind management of my thesis committee and the department of physics. I only regret not benefiting even more from his knowledge.

I would like to thank Mahdi for assisting me during the hardest times of my life. Due to my health conditions, I was forced to quit graduate school, that is when Mahdi stepped in and offered help to me. I can honestly say without his intervention, this thesis would not exist. I will never forget his help. I owe a debt of gratitude for both his constant academic and non-academic support.

I would also like to thank Joel for his strong sense of responsibility which is considerably greater than his academic duty requires from him; for giving me intellectual freedom in my work while simultaneously not sacrificing his high academic standards; for the time he devoted to me, the feedback, and his precise and comprehensive emails. All of his support assured me that he was always there to aid my academic future.

Other than my advisors, I had the great chance to learn from Bryan Roberts through email and reading his excellent PhD dissertation and papers very closely related to my thesis topic. His works and guidance helped me a lot, and so I am deeply grateful.

I would like to express my deep gratitude to Howard Curzer for showing me the humane aspect of a philosophy teacher. I greatly benefited from his philosophical insights during our valuable conversations and exciting meetings. Without a doubt he is the nicest person I have ever met.

I would be remiss if I did not thank Daniel Nathan who supported me kindly like a father during my endeavors. I am not the only student he has helped prevent from slipping through the cracks, so I am thankful to him for being there for both me and all the others.

Finally, I would like to thank Stephen Buchok, my psychiatrist. Although he unfortunately passed away more than one year ago, I will remember him as my dutiful and kind-hearted doctor. Also I would like to thank Debrajean Wheeler and Joyce Norton, kind staff members of the Philosophy and Physics departments, respectively.

# TABLE OF CONTENTS

# ABSTRACT

Symmetries have a crucial role in today's physics. In this thesis, we are mostly concerned with time reversal invariance ($T$-symmetry). A physical system is time reversal invariant if its underlying laws are not sensitive to the direction of time. There are various accounts of time reversal transformation resulting in different views on whether or not a given theory in physics is time reversal invariant. With a focus on quantum mechanics, I describe the standard account of time reversal and compare it with my alternative account, arguing why it deserves serious attention. Then, I review three known ways to $T$-violation in quantum mechanics, and explain two unique experiments made to detect it in the neutral $K$ and $B$ mesons.

# LIST OF FIGURES

# CHAPTER I

# INTRODUCTION

The nature of time is one of the central questions of physics and philosophy, especially metaphysics and philosophy of science. Many controversial issues as direction of time, reversibility, backward causation and time travel are related to this topic. In the past, physics and philosophy were not sharply distinct disciplines, but nowadays, modern physicists and analytic philosophers do not have the same approach toward these questions. Nevertheless, both physicists and philosophers agree that these crucial questions are indispensable to our ultimate understanding of nature and the shaping of our worldview.

## 1.1. Spatiotemporal Symmetries

Symmetry is also critical to the structure of our best theories in physics. In at least two ways, various kinds of symmetries or invariances have a very important role in fundamental physics, especially quantum theory and relativity:

> First, we may attribute specific symmetry properties to phenomena or to laws (*symmetry principles*). …Second, we may derive specific consequences with regard to particular physical situations or phenomena on the basis of their symmetry properties (*symmetry arguments*) [1].

Symmetry under time reversal or time reversal invariance is a kind of symmetry principle, which is the main focus of this thesis. Space inversion (P or parity) and *CPT* symmetry (or *CPT* theorem for Charge conjugation, Parity, and Time invariance) are closely related to our discussion. Below, we introduce parity $(P)$[1] and time reversal $(T)$ transformations along with other spatiotemporal symmetries. $C$ is called charge conjugation transformation, and replaces a particle with its antiparticle. $CPT$ symmetry says that in specific (but general enough) conditions, quantum phenomena are invariant under combination of $C$, $P$ and $T$ transformations [2]. Consider any possible combination of these transformations, like $P$ itself or $CP$ denoted by $X$, and a given state of the quantum system. If the state is invariant under $X$, it is said that the state is $X$-even, otherwise it is said that the state is $X$-odd. Employing this convention, $CPT$ symmetry states that under certain restrictions, every quantum state is $CPT$-even [2].

As is depicted in figure 1, there are three spatial and two temporal symmetries. They have a very special role in the foundation of physics, let's see why and how. Space translation symmetry means that the location of the origin of a coordinate system is arbitrary. In other words, no physical system is sensitive to its location in space, or the behavior of the system is the same no matter where the coordinate origin is located. This is what is meant by uniformity of space, and results in the conservation of linear momentum. Also, the orientation of the axes is arbitrary. This property is called isotropy of space, and results in the conservation of angular momentum [2].

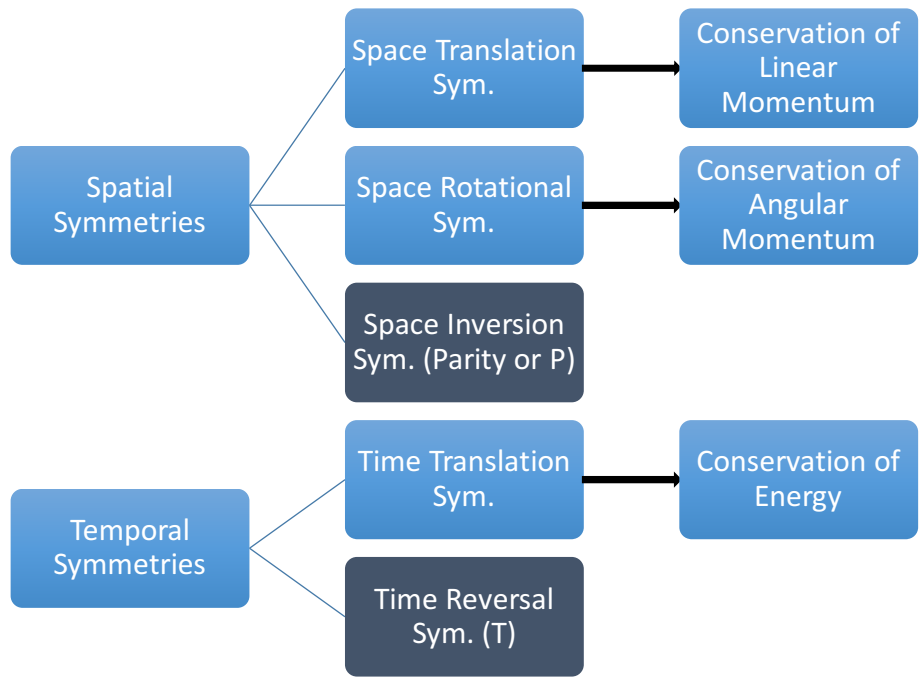

*Figure 1. Spatial and temporal symmetries.*

---

[1] I use $X$ to denote both $X$ symmetry and $X$ transformation.

The other fundamental symmetry is time translation symmetry, as a result of uniformity of time. This means that choice of time origin, $t = 0$, is conventional. That is to say, all dynamics are invariant under time translation, or change of time origin. This symmetry yields conservation of energy [2]. These symmetries, and the corresponding conservation laws are never experimentally violated. It seems that they are built in the constitution of nature, otherwise, we could not be assured that the same two experiments would have same results whenever and wherever are they carried out. In other words, it is by virtue of the uniformity of space and uniformity of time that we can explain and predict phenomena in science, and this is a very important role that these symmetries play in the foundation of science.

However, the situation is somewhat different for space inversion symmetry and time reversal symmetry. Firstly, unlike translation and rotation which may be described in terms of continuously varying sets of parameters, parity and time reversal transformations cannot be carried out continuously. For this reason, the former are called "proper" transformations while the latter are called "improper" transformations. Secondly, for a long time, it was thought that $P$ and $T$ symmetries hold in nature too, but in the second half of the 20<sup>th</sup> century, they were experimentally shown to be violated [2]. Space inversion is interchanging the right-handed and left-handed coordinate systems, or negating all space coordinates (point reflection). For example, in three dimensions:

$$\mathbf{r} = \begin{pmatrix} x \\ y \\ z \end{pmatrix} \xrightarrow{P} -\mathbf{r} = \begin{pmatrix} -x \\ -y \\ -z \end{pmatrix}.$$

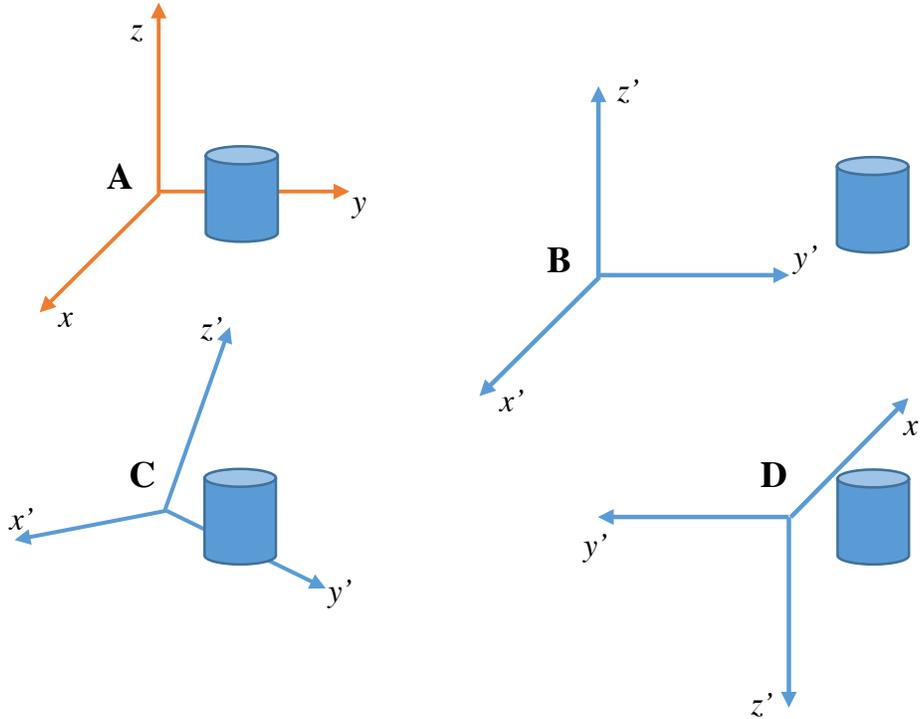

*Figure 2. Spatial symmetry transformations: A) Original coordinate origin (Right-handed) B) Translated coordinate origin, C) Rotated coordinate origin, D) Inversed coordinate origin (Left-handed)*

We cannot construct parity merely via rotation, a mixture of mirror (plane) reflection *and* $\pi$ rotation is needed, so $P$-violation does not violate isotropy of space:

$$\mathbf{r} = \begin{pmatrix} x \\ y \\ z \end{pmatrix} \xrightarrow[\substack{x-z \text{ plane}}]{\substack{Step\ 1: \\ mirror \\ reflection}} \begin{pmatrix} x \\ -y \\ z \end{pmatrix} \xrightarrow[\substack{\text{around y}}]{\substack{Step\ 2: \\ \pi \text{ rotation}}} \begin{pmatrix} -x \\ -y \\ -z \end{pmatrix} = -\mathbf{r}$$

In 1957, Chien Shiung Wu's team observed $P$-violation in the $\beta$-decay of $^{60}_{27}\mathrm{Co}$ ($\beta - \mathrm{Co}$). Actually, they detected $P$-violation in step 1 above, that is to say, they found that an experiment which is set up like the mirror image of the original $\beta - \mathrm{Co}$ experiment does

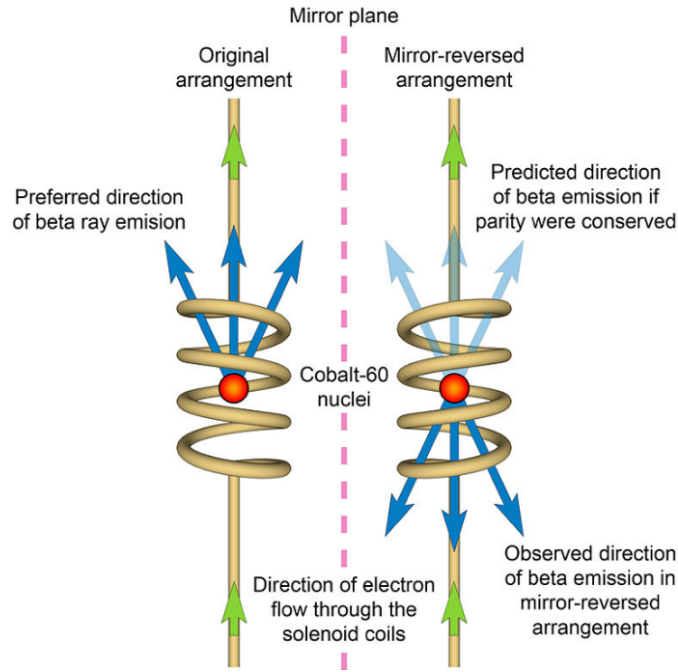

*Figure 3*. *Wu's experiment basic idea.*

*not* behave like the mirror image of the original $\beta - \text{Co}$ experiment [3]. The discovery of *P*-violation was a significant contribution to particle physics and the development of the standard model [2]. For their role in this discovery, Wu was awarded Wolf Prize in 1978 and before that, her colleagues, Tsung Dao Lee and Chen Ning Yang were awarded the Nobel Prize in Physics in 1957.

## 1.2. Time Reversal Invariance

What is meant by time reversal invariance? Commonly, a physical process is said to be time reversal invariant if its reverse is also allowed by laws of nature. However, there is a great deal of controversy regarding the exact physical meaning of the reverse of a process. According to the simplest view, to reverse a process it suffices to reverse the order of the instantaneous states of the process. That is what exactly happens when we play the film of a macroscopic process in reverse, for example the collision of two billiard balls. In this case, both what the film and the film played in reverse show are allowed by laws of classical mechanics, and by just watching the films we cannot say which is the real process and which is its reverse. Each is the reverse of the other, and both are equally plausible under normal conditions [4].

Suppose a finite process happening between $t_0$ and $t_N$ as a sequence of instantaneous states $S(t_0)$, $S(t_1)$, $S(t_2)$, ... , $S(t_N)$. The above view implies that the reverse of this process is just the sequence of instantaneous states $S(t_N)$, ... , $S(t_2)$, $S(t_1)$, $S(t_0)$ happening in the same interval of time. To put it more formally, a finite process of $S(t_0)$, $S(t_1)$, $S(t_2)$, ... , $S(t_N)$ is time reversal invariant if the reversed process of $S(t_N)$, ... , $S(t_2)$, $S(t_1)$, $S(t_0)$ is also allowed by the laws of nature. We can generalize this definition to be applied to a theory. A theory is time reversal invariant if for any (in)finite sequence of instantaneous states ... , $S(t_0)$, $S(t_1)$, $S(t_2)$,... allowed by the theory, the reverse sequence of time-reversed states, ... ,$S(t_2)$, $S(t_1)$, $S(t_0)$, ... is also allowed [4].

Recalling the billiard balls case, it seems that according to this definition, classical mechanics theory is time reversal invariant. But the situation is more complicated for the electromagnetism and quantum theories [4]. The status of time reversal in these theories is discussed in the next chapter. Historically, it was thought that time reversal symmetry must not be violated in the microscopic theories underlying classical mechanics as a time reversal symmetric theory [2]. However, as will be explained in chapter 3, in 1964, James Cronin, Val Fitch and their coworkers detected indirect evidence for $T$-violation in the decay of $K$ mesons and won the Physics Nobel Prize in 1980 [5]. After that, two direct observations of $T$-violation were made in the decay of $K$ and $B$ mesons in 1998 and 2012 by CPLEAR and BaBar collaboration, respectively [6, 7].

## 1.3. Temporal Asymmetry of Thermodynamics

Contrary to the seemingly obvious symmetry of classical mechanics under time reversal, the other macroscopic theory in physics, namely thermodynamics, shows completely asymmetric behavior. The thermodynamics time asymmetry is one of the most conspicuous properties of nature. There are infinite examples of temporal asymmetry in thermodynamics: heat flows from hot to cold, never the reverse. The smell of perfume spreads throughout its environment, never the reverse. Airplane jet engines convert fuel energy into work and thermal energy, never the reverse. Thermodynamics is able to explain these asymmetric phenomena as a result of its assertion that systems automatically evolve to equilibrium states in the flow of time, but do not automatically evolve away from equilibrium states [8].

There is a big puzzle here: how thermodynamics, as a non-fundamental theory, can be asymmetric if it is thought that fundamental laws underlying it are time symmetric? The common view is that the asymmetry of thermodynamics must be reduced to either asymmetric initial conditions or asymmetric underlying laws. Laws of classical mechanics are not a good candidate, because classical mechanics has time symmetric dynamics. So, many thinkers try to base the asymmetry of thermodynamics in electromagnetism or quantum mechanics. But there are many difficulties. The remaining option is appealing to asymmetric initial conditions to solve the problem [8].

Assuming temporally asymmetric boundary conditions makes it possible to have a world evolving toward equilibrium but not evolving away from it. For instance, a cosmological hypothesis, as David Albert calls it the "Past Hypothesis", states that in the very far past entropy was extremely lower than now. He claims that earlier states had lower entropy than current ones because, according to the Past Hypothesis, the universe started in a super small part of its available phase space [4]. Although it seems that the Past Hypothesis solves this puzzle, there are some concerns about it which are irrelevant to our discussion here.

## 1.4. Quantum Mechanics

Quantum mechanics as a microscopic theory describing the physical reality is much more complex than classical mechanics and thermodynamics. Unlike classical mechanics, we cannot easily imagine if it is invariant under time reversal, and also contrary to thermodynamics, we cannot observe its temporal asymmetric behavior in everyday life. But we are very curious to know what really happens in quantum mechanics under time reversal transformation. A reason is that quantum mechanics is our most fundamental theory in physics, underlying other microscopic and macroscopic theories like statistical mechanics, electromagnetism, classical mechanics and thermodynamics. It seems that we can have a clearer and more detailed picture of time reversal in these theories if we understand it at the quantum level [4, 9].

Although physicists may not be very interested, philosophers like to have a precise definition of quantum mechanics as a queer but successful theory. Like other philosophical questions there is not a unique and undoubtable answer to it. But we can summarize:

> Quantum mechanics is, at least at first glance and at least in part, a mathematical machine for predicting the behaviors of microscopic particles — or, at least, of the measuring instruments we use to explore those behaviors — and in that capacity, it is spectacularly successful: in terms of power and precision, head and shoulders above any theory we have ever had. Mathematically, the theory is well understood; we know what its parts are, how they are put together, and why, in the mechanical sense (i.e., in a sense that can be answered by describing the internal grinding of gear against gear), the whole thing performs the way it does, how the information that gets fed in at one end is converted into what comes out the other. The question of what kind of a world it describes, however, is controversial; there is very little agreement, among physicists and among philosophers, about what the world *is like* according to quantum mechanics [10].

It should be mentioned that here "quantum mechanics" is understood in its broad sense, something like "quantum theory", including particle physics and quantum field theory, in addition to what is commonly thought as "quantum mechanics", in the narrow sense, in the elementary quantum mechanics text books[2]. And this is another reason why contrary to classical mechanics or electromagnetism, we cannot consider it as a definite body of concrete laws or a set of limited subject matter — it has a very wide scope. Quantum mechanics is "microscopic" in the sense that it is describing the micro structure of the universe, and it is "fundamental" in the sense that it cannot be reduced to any other theory dealing with physical reality [11]. However, the ambiguity in the quantum mechanics definition neither prevents us from studying time reversal symmetry, as the main subject of our interest here, nor decreases the importance of this question.

## 1.5. The Problem and its Significance

Are quantum phenomena time reversal invariant or not? Why and how? These are the questions we are trying to answer in this thesis. By "quantum phenomena", we mean all quantum processes or transitions which are taken into account in the quantum mechanics discussed above. In other words, for a phenomenon to act according to quantum mechanics means that it can be described or explained within the context of quantum mechanics. Obviously, these definitions suffer from ambiguity in more or less the same way the quantum mechanics definition does. As was mentioned before in the definition of time reversal invariance for the processes and theories, if all processes admitted by a theory, in our case quantum mechanics, are time reversal invariant, then the theory is said to be time reversal invariant.

---

[2] As an illustrative example of elementary quantum mechanics text book, please see *Introduction to Quantum Mechanics* by David Griffiths (1982).

But what is the real importance of this question and why should we care about it? In the beginning, I mentioned some well-known problems related to the question of the nature of time and its role in the physics, like problems of the direction of time, reversibility, backward causation and time travel. They are closely related to our question. As an example, below I will briefly review its application in the problem of direction of time as is discussed by John Earman, then I will mention some of its other applications.

Earman believes spacetime must be locally temporally orientable by arguing that "for any point $p \in \mathcal{M}$ a small enough neighbourhood $N(p)$ can be chosen that $N$ is simply connected, and any simply connected manifold with a Lorentz signature metric admits a continuous non-vanishing timelike vector field" [12]. But he notices that this does not rule out the possibility of some weird multiple connectivity in the large scale level that prohibits us from making a globally consistent distinction between past and future. The induction from local orientability to global orientability does not work because the former can hold everywhere without implying the latter [12]. But he claims that a more sophisticated kind of induction can help us to derive global orientability:

> *If* the laws of physics are "universal" in the sense that they are the same in every region of spacetime, *if* by local investigations we manage to find (some of) the basic laws of physics, and *if* these laws are not time reversal invariant, then we can infer that in our universe there is a globally consistent distinction between past and future. Roughly the idea is this. Choose any closed path in the spacetime. Suppose for purposes of *reductio* that the transport of a timelike vector around some such path by some method that is continuous and keeps timelike vectors timelike results in a flip in time sense when the vector returns to the starting point. Along this path choose a chain of overlapping simply connected neighbourhoods. Use the failure of time reversal invariance of the laws to pick out the future direction of time in each of these neighbourhoods. … Thus the future direction picked out in two adjacent neighbourhoods *N* and *N'* must agree in the overlap *N∩N'*. But by the *reductio* assumption this agreement must fail when the chain of neighbourhoods closes. Since a contradiction has been reached the *reductio* assumption must be false and the spacetime is globally temporally orientable [12].

He claims that the above argument makes us certain that our universe is temporally orientable. Furthermore, he claims that our universe is actually temporally oriented by adding the fact that "a temporal orientation is needed to sort the dynamically possible from the dynamically impossible histories when time reversal invariance fails" [12]. This was a summary of Earman's argument in the favor of the direct role of time reversal non-invariance in the problem of direction of time. Below I quote some of its other applications by Robert Sachs [2]:

> As far as is known at present, the electromagnetic and strong interactions responsible for the structure and general dynamic behavior of atoms and atomic nuclei are invariant under time reversal. This invariance has important consequences for the properties of stationary states, scattering and reaction amplitudes, and (electromagnetic) radiative transitions of such systems. … they may be used to test the assumption that these or other interactions are in fact $T$ invariant.

Usually in quantum mechanics there are associated with the invariance of Hamiltonian a conservation law and some degree of degeneracy of the energy states. Invariance of the Hamiltonian under $T$ has different implications. Because $T$ is anti-unitary rather than unitary, it is not directly related to a Hermitian observable, and the invariance does not lead to a conservation law. There is an implication of twofold degeneracy (Keramer's Degeneracy) for "odd" systems … and there are additional implications for the stationary states of any multiparticle system. The latter may be expressed as reality conditions on the wave functions…

We discussed some necessary and prerequisite points about time reversal invariance in this introductory chapter. In the next chapter, the standard account of time reversal invariance in the classical mechanics, electromagnetism and quantum mechanics will be discussed, with the focus on quantum mechanics. Also, I will describe my own account of time reversal invariance comparing it with the standard one, and argue why my account deserves attention. In chapter 3, I will review three known ways yielding to $T$-violation in quantum mechanics, and then in chapter 4, I will explain two important experiments made to detect $T$-violation in the neutral $K$ and $B$ mesons. In the conclusion, I highlight the major points emerging form this thesis.

# CHAPTER II

# TIME REVERSAL INVARIANCE

I begin this chapter by exploring the common understanding of time reversal symmetry in the classical mechanics, electromagnetism and especially quantum mechanics. Then, I will discuss the general concept of time reversal invariance and its two main accounts, namely the standard account and Albert's account. I will present and examine my own alternative account of time reversal at the end. Also based on the invariance of the Schrödinger equation under time reversal transformation specific to quantum mechanics, I will briefly examine time reversal in the three main interpretations of quantum mechanics.

## 2.1. Time Reversal Invariance in Classical Mechanics

What is a time reversal transformation? Assume we film a pendulum clock in action. If the film is played in reverse, the result will be a new "reversed" motion of clock's hands and pendulum. This is what we intuitively think of as a time reversal transformation. But how is this transformation described mathematically? In the Newtonian formulation of classical mechanics, it is merely the reversal of the order of events in a trajectory $\mathbf{x}(t)$. In other words, if $\mathbf{x}(t)$ is the curve depicting the position of the pendulum over time, then the time-reversed trajectory is given by $\mathbf{x}(-t)$ [13].

We can easily see the requirement that $\mathbf{F}(\mathbf{x}; t) = \mathbf{F}(\mathbf{x}; -t)$ to guarantee time reversal invariance in Newtonian mechanics is equivalent to requiring $\mathbf{x}(-t)$ to satisfy Newton's equation whenever $\mathbf{x}(t)$ does. However, in the Hamiltonian formulation, where $\mathbf{q}$ denotes the position and $\mathbf{p}$ denotes the momentum of a given particle, reversing the order of events in a trajectory $(\mathbf{q}(t), \mathbf{p}(t))$ is not enough. Besides that, it needs to reverse momentum while preserving position: $T(\mathbf{q}, \mathbf{p}) = (\mathbf{q}, -\mathbf{p})$. Here, the requirement that $H(\mathbf{q}, \mathbf{p}) = H(\mathbf{q}, -\mathbf{p}) + k$ (for some $k \in \mathbb{R}$) to guarantee time reversal invariance in Hamiltonian mechanics is the same as requiring $(\mathbf{q}(-t), -\mathbf{p}(-t))$ to satisfy Hamilton's equations whenever $(\mathbf{q}(t), \mathbf{p}(t))$ does [13].

### 2.1.1. When Classical Mechanics Is Time Reversal Invariant?

We have seen the naïve claim that classical mechanics is time reversal invariant many times in the elementary textbooks on the classical mechanics. But it is easy to find a counterexample: a classical system with a so-called "dissipative" force. For example, Newton's laws, as well as Hamilton's equations, allow trajectories in which a mass moves along a smooth surface, suffering from the friction force, until finally stops [13]. However, we know that the time-reversed trajectory of a mass suddenly accelerating from rest is not a possible solution to the Newton's laws or Hamilton's equations. Thus we need to limit our claim scope: "Classical mechanical systems that are 'conservative' are also time reversal invariant" [13].

Here, the truth of this claim depends on the exact definition of the term "conservative." Below, two common definitions are considered, and it is shown that they are not sufficient to guarantee time reversal invariance of the system under description. Finally, a sufficient definition is suggested.

"No free work" definition of a conservative system: this is a common textbook definition of a conservative system in the Newtonian formulation. This definition employs the quantity of work required to move a system between two points 1 and 2:

$$W_{12} = \int_1^2 \mathbf{F}.\,d\mathbf{x}$$

According to this definition, if the trajectory between points 1 and 2 is a closed loop (i.e. $W = \oint \mathbf{F}.\,d\mathbf{x}$), then $W = \mathbf{0}$. Thus, a conservative system does not admit "free work", or if a process ends in the exact starting state, then total work done is zero. If the force field is such that the work $W_{12}$ is equal for any path between points 1 and 2, then the force and the system is called to be conservative in this sense [13].

"$dH/dt = 0$" definition of a conservative system: this is another standard definition of a "conservative" system, but this time in the Hamiltonian mechanics. Here, the Hamiltonian $H$ is often interpreted as total energy of a system. In principle, "conservative" implies that $H$ is a conserved quantity, or $dH/dt = 0$ [13].

However, the first definition is not sufficient to guarantee time reversal invariance. This is a simple example: take a particle in 3-dimensional space, with position $\mathbf{x}$ and velocity $\dot{\mathbf{x}}$. Suppose the particle is subject to a force field defined by,

$$\mathbf{F} = \mathbf{x} \times \dot{\mathbf{x}}$$

that is, the force on the particle is orthogonal to both its position and velocity vectors. This system is "conservative" in the first sense. The reason is that $\mathbf{F}$, the cross product of $\mathbf{x}$ and $\dot{\mathbf{x}}$, is orthogonal to $\mathbf{x}$, and hence to $d\mathbf{x}$. So, the line integral characterizing work $W_P$ along any path $P$ is zero [13].

Nevertheless, the system is not time reversal invariant. This is because the system moves in a preferred direction, i.e. the direction orthogonal to $\mathbf{x}$ and $\dot{\mathbf{x}}$ given by the right hand rule. But under time reversal, the velocity vector is reversed and so that referred direction is not preserved. To check this formally, we can see that $\mathbf{F}(\mathbf{x}, -t) = \mathbf{x} \times (-\dot{\mathbf{x}}) = -\mathbf{F}(\mathbf{x}, t)$. So $\mathbf{F}(\mathbf{x}, -t) \neq \mathbf{F}(\mathbf{x}, t)$, and time reversal invariance fails. Thus, being conservative in this sense is not sufficient for time reversal invariance [13].

Also, the second definition may fail to result in time reversal invariance. There are many conservative systems of this type that break time reversal invariance. For example, consider a particle described by the somewhat unphysical Hamiltonian $H = \|\mathbf{p}\|$. It can be shown that for this Hamiltonian, $dH/dt = \partial H/\partial t = 0$, therefore this system is conservative in the required sense. However, since $H(\mathbf{q}, -\mathbf{p}) \neq H(\mathbf{q}, \mathbf{p}) + k$, the system is not time reversal invariant [13].

Thus we have seen that there are various ways in which a system that is conservative in the broad sense of "conserving energy" violates time reversal invariance. To guarantee time reversal invariance, a stronger condition is needed. In the context of Newtonian mechanics, some people define a conservative system to be one in which all forces have a particular functional form:

$$\mathbf{F} = -\nabla V(\mathbf{x})$$

that is, a force which can be expressed as the gradient of a time-independent potential $V$. In this definition, Newton's equation is obviously time reversal invariant, because the right hand side of the above formula has no time-dependence, and consequently $\mathbf{F}(\mathbf{x}, t) = \mathbf{F}(\mathbf{x}, -t)$ [2, 13].

## 2.2. Time Reversal Invariance in Electromagnetism

To study time reversal in electromagnetism we ask how the quantities involved in the Maxwell's equations transform under time reversal, or replacing $t$ with $-t$. As we have seen in the previous section, the time reverse of a particle moving from point 1 to 2 is a particle moving from point 2 to 1, implying that its velocity $\mathbf{v}$ must flip sign under time reversal. We know that $\mathbf{J} = \rho\mathbf{v}$, and that the charge density $\rho$ is invariant under time reversal, because it is not supposed to be time dependent, neither directly like $\mathbf{v} = \frac{d\mathbf{x}}{dt}$, nor indirectly like spin. Consequently the electric current density $\mathbf{J}$ will flip sign under time reversal as well [14]. So now let's look at Maxwell's equations in Gaussian units convention:

$$\nabla \cdot \mathbf{E} = 4\pi\rho$$

$$\nabla \times \mathbf{E} = -\frac{1}{c}\frac{\partial \mathbf{B}}{\partial t}$$

$$\nabla \cdot \mathbf{B} = 0$$

$$\nabla \times \mathbf{B} = \frac{1}{c}\left(4\pi\mathbf{J} + \frac{\partial \mathbf{E}}{\partial t}\right)$$

How do electric field $\mathbf{E}$ and magnetic field $\mathbf{B}$ change under time reversal? Here, we can see that Maxwell's equations do not (uniquely) determine the sign of $\mathbf{E}$ and $\mathbf{B}$ after replacing $t$ with $-t$. However, the common assumption among physicists is that electromagnetism is invariant under time reversal. Given this assumption, in addition to the above points that $\mathbf{J}$ flips sign under time reversal but $\rho$ does not change, Maxwell's equations imply that the electric field $\mathbf{E}$ is invariant under time reversal, while the magnetic field $\mathbf{B}$ negates [14]. So given the common assumption of the time reversal invariance of the electromagnetism, the related quantities transform under time reversal as below:

$$\mathbf{v} \xrightarrow{\text{T}} -\mathbf{v}$$

$$\mathbf{J} \xrightarrow{\text{T}} -\mathbf{J}$$

$$\rho \xrightarrow{T} \rho$$

$$\mathbf{E} \xrightarrow{T} \mathbf{E}$$

$$\mathbf{B} \xrightarrow{T} -\mathbf{B}$$

## 2.3. Time Reversal Invariance in Quantum Mechanics

To begin the discussion of time reversal in quantum mechanics, most of the textbooks assume three myths [15]:

> **Myth 1.** The preservation of transition probabilities ($|\langle T\psi, T\phi \rangle| = |\langle \psi, \phi \rangle|$) is a definitional feature of time reversal, with no further physical or mathematical justification …
>
> **Myth 2.** The anti-unitary character of time reversal can only be established by fiat, or by appeal to particular transformation rules for 'position' and 'momentum' …
>
> **Myth 3.** The way that position and momentum transform under time reversal can only be justified by appeal to their classical analogues …

However, Bryan Roberts tries to dispel these myths by presenting his own reasoned three-step plan which is aimed at bringing about the very results which these myths do in a dogmatic manner. Let's see a summary of Roberts' three-step plan to construct $T$ in a clear and systematic way: in the first step, he employs the fact that the direction of time is irrelevant to the question of whether or not two states are orthogonal. As he points out, this fact implies that $T$ should be unitary or anti-unitary. A unitary operator is a bijection $T \colon \mathcal{H} \to \mathcal{H}$ that satisfies these two conditions:

1) (adjoint inverse) $T^*T = TT^* = I$.

2) (linearity) $T(a\psi + b\phi) = aT\psi + aT\phi$ for any $\psi, \phi \in \mathcal{H}$.

And an anti-unitary operator is a bijection $T \colon \mathcal{H} \to \mathcal{H}$ that satisfies these two conditions:

1) (adjoint inverse) $T^*T = TT^* = I$.

2) (anti-linearity) $T(a\psi + b\phi) = a^*T\psi + b^*T\phi$ for any $\psi, \phi \in \mathcal{H}$.

The second step assumes that there exists at least one physically plausible non-trivial system that is time reversal invariant. As he establishes, this assumption yields the result that $T$ is anti-unitary. The third step is based on the reasonable assumption that the meaning of $T$ is independent of translations or rotations in space, representing two kinds of familiar symmetry transformations. He uses this assumption to derive aforementioned transformation rules for desired observables, i.e. position, momentum and spin [15].

### 2.3.1. First Step

His first step is based on the Uhlhorn's and Wigner's theorems (see below). Assuming Uhlhorn's theorem, Roberts uses Wigner's theorem to conclude that transformations like time reversal $T$ (and rotation, translation, etc.), can always be represented either by a unitary operator or by an anti-unitary one [15].

> **Uhlhorn Theorem**. *Let* **T** *be any bijection on the ray space* $\mathfrak{Y}$ *of a separable Hilbert space* $\mathcal{H}$ *with dimension greater than 2. Suppose that* $\Psi \perp \Phi$ *if and only if* **TΨ** $\perp$ **TΦ**. *Then,*
>
> $$\langle \mathbf{T}\Psi, \mathbf{T}\Phi \rangle = \langle \Psi, \Phi \rangle.$$
>
> *Moreover, there exists a unique (up to a constant)* $T: \mathcal{H} \to \mathcal{H}$ *that implements* **T** *on* $\mathcal{H}$, *in the sense that* $\psi \in \Psi$ *iff* $T\psi \in \mathbf{T}\Psi$, *and which satisfies* $(|\langle T\psi, T\phi \rangle| = |\langle \psi, \phi \rangle|)$ *for all* $\psi, \phi \in \mathcal{H}$ [13].

As Roberts says,

---

[3] In [13], [15], and [18], Roberts does not denote vector and scalar variables differently. He uses italic (non-bold) letters for both. So in the direct quotes from him, bold letters do not denote vectors.

The intuition underlying Uhlhorn's condition is that if we film a physical system that allows these two properties, then the two propositions remain mutually exclusive *no matter whether the film is playing forward or in reverse*. The facts about mutual exclusivity should be independent of anything to do with the facts about the direction of time [13].

**Wigner Theorem**. For any **T** satisfying the Uhlhorn Theorem, there exists a Hilbert space operator $T$ that implements it which is either unitary or anti-unitary [16].

### 2.3.2. Second Step

Firstly, Roberts argues,

The time-reversing transformations can be minimally identified as bijections on the set of trajectories $\psi(t) = e^{-itH}\psi$ that take the form,

$$\psi(t) \rightarrow T\psi(-t) = Te^{itH}\psi,$$

where $T$ at this point is an arbitrary unitary or anti-unitary operator, possibly even the identity operator [13].

Then, in the next move which is a little bit tricky, he says that if $T$ is a true symmetry transformation, the equation $\psi(t) = e^{-itH}\psi$ must be valid for any transformed trajectory too. If we denote the transformed trajectories $T\psi(-t) = Te^{itH}\psi$ by $\phi(t)$, we must have the same unitary law, $\phi(t) = e^{-itH}\phi$. Now by replacing $\phi(t)$ with $Te^{itH}\psi$ in the left side of the last equation, we have $Te^{itH}\psi = e^{-itH}\phi$. And recalling $(t) \coloneqq T\psi(-t) = Te^{itH}\psi$, we conclude that $\phi = T\psi$, and so $Te^{itH}\psi = e^{-itH}\phi = e^{-itH}T\psi$. Thus finally we have:

$$Te^{itH}\psi = e^{-itH}T\psi$$

for all $\psi$. However, this equation does not apply to all Hamiltonians, e.g. the Hamiltonians describing the weak interactions [13, 15]. But according to Roberts, "…we do suppose that *at least one* of these Hamiltonians—perhaps a particularly simple one with no

interactions—is $T$-reversal invariant. This turns out to be enough to establish that $T$ is anti-unitary" [15]. This is done by employing the last equation in the below proposition.

**Proportion 1**. *Let $T$ be a unitary or anti-unitary bijection on a separable Hilbert space $\mathcal{H}$. Suppose there exists at least one density-defined self-adjoint operator $H$ on $\mathcal{H}$ that satisfies the following conditions.*

- (i)    (positive)  $0 \leq \langle \psi, H\psi \rangle$ *for all $\psi$ in the domain of $H$.*
- (ii)   (non-trivial) $H$ *is not the zero operator.*
- (iii)  (T-reversal invariant)  $Te^{itH}\psi = e^{-itH}T\psi$ *for all $\psi$.*

*Then $T$ is anti-unitary.*

**Proof.**  Condition  (iii)  implies  that  $e^{itH} = Te^{-itH}T^{-1} = e^{T(-itH)T^{-1}}$. Moreover, Stone's theorem guarantees the generator of unitary group $e^{itH}$ is unique when $H$ is self-adjoint. So $itH = -TitHT^{-1}$. Now, suppose for reductio that T is unitary, and hence linear. Then we can conclude from the above that $itH = -itTHT^{-1}$, and hence $THT^{-1} = -H$. Since unitary operators preserve inner    products,    this    gives,    $\langle \psi, H\psi \rangle = \langle T\psi, TH\psi \rangle = -\langle T\psi, HT\psi \rangle$. But condition (i) implies both $\langle \psi, H\psi \rangle$ *and* $\langle T\psi, HT\psi \rangle$ are non-negative so we have,

$$0 \leq \langle \psi, H\psi \rangle = -\langle T\psi, HT\psi \rangle \leq 0.$$

It follows that $\langle \psi, H\psi \rangle = 0$ for all $\psi$ in the domain of $H$. Since $H$ is defined, this is only possible if $H$ is the zero operator, contradicting Condition (ii). Therefore, since $T$ is not unitary, it can only be anti-unitary. [15]

The inference is based on the assumption that there is at least one possible dynamical system—not necessarily even a realized one—that is time reversal invariant. If there is such a system, then time reversal can only be anti-unitary [15].

### 2.3.3. Third Step

According to Roberts, the anti-unitarity of time reversal and the canonical commutation relations are not enough to guarantee these transformation rules:

$$\mathbf{Q} \rightarrow \mathbf{Q}$$

$$\mathbf{P} \rightarrow -\mathbf{P}$$

$$\boldsymbol{\sigma} \rightarrow -\boldsymbol{\sigma}$$

where $\mathbf{Q}$, $\mathbf{P}$ and $\boldsymbol{\sigma}$ are position, momentum and spin observables, respectively. To satisfy these conditions, Roberts just relies on the general and plausible assumptions that space is homogeneous and isotropic under time reversal. He derives the first two transformation rules from the former and the last one from the latter [15]. Suppose that $U_a$, $V_b$, and $R_\theta$ are generators of spatial translations, boosts in velocity and spatial rotation, respectively. Now let's explain the above assumptions. Consider these three propositions [15]:

a) "If we first time reverse a state and then translate it, the result is the same as when we first translate and then time reverse", or $U_a T \psi = T U_a \psi$.

b) "If we time reverse a system and then apply a boost in velocity, then this is the same as if we had boosted in the opposite spatial direction and then applied time reversal", or $V_b T \psi = T V_{-b} \psi$.

c) If we first time reverse a system and then rotate it, then this is the same as if we first rotate it and then time reverse, or $R_\theta T \psi = T R_\theta \psi$.

Space is homogeneous under time reversal in the sense that propositions (or equations) a *and* b both hold. Space is isotropic under time reversal in the sense that proposition (or equation) c holds. I neither want to go into the details of the mathematical structure of the three generators mentioned above, nor the mathematics of the equations a, b, and c *after* applying those details. I just mention the final results of each of the above cases by referring to its corresponding uppercase letter [15]:

A) $T\mathbf{Q}T^{-1} = \mathbf{Q}$ or $\mathbf{Q} \xrightarrow{T} \mathbf{Q}$.

B) $T\mathbf{P}T^{-1} = -\mathbf{P}$ or $\mathbf{P} \xrightarrow{T} -\mathbf{P}$.

C) $T\boldsymbol{\sigma}T^{-1} = -\boldsymbol{\sigma}$ or $\boldsymbol{\sigma} \xrightarrow{T} -\boldsymbol{\sigma}$.

He derives these transformations rules without appealing to anything other than the assumption of homogeneity and isotropy of space under time reversal. It can be easily shown that in the Schrödinger wavefuncation representation in which $\mathbf{Q}\psi(x) = x\psi(x)$ and $\mathbf{P}\psi(x) = i\frac{d}{dx}\psi(x)$, the time reversal transformation,

$$T = K$$

implements A and B transformation rules, where $K$ is the conjugation operator, $K\psi(x) = \psi^*(x)$. As Roberts proves, $T$ is *unique* up to a constant, i.e. it is the only way we can define $T$, not just a possible way [15].

In addition, Roberts proves that there is a *unique $T$* (up to a constant),

$$T = \sigma_2 K$$

which implements the transformation rule C for half-spin particles. Here, $\sigma_2$ is the second Pauli matrix $\begin{pmatrix} 0 & -i \\ i & 0 \end{pmatrix}$, and $K$ is the conjugation operator mapping $\psi = \alpha\begin{pmatrix} 1 \\ 0 \end{pmatrix} + \beta\begin{pmatrix} 0 \\ 1 \end{pmatrix}$ to its conjugate $\psi^* = \alpha^*\begin{pmatrix} 1 \\ 0 \end{pmatrix} + \beta^*\begin{pmatrix} 0 \\ 1 \end{pmatrix}$. To double-check, we can easily see that both above $T$s are anti-unitary, as expected [15]. This was a very short summary of Roberts' three-step plan for constructing $T$, the time reversal operator in quantum mechanics.

## 2.4. Discussion

In this section, first, I will discuss two well-known accounts of time reversal, namely the standard account and Albert's account, and then I will present my own alternative account and compare it with the standard one. At the end, I will examine time reversal invariance of three main interpretations of quantum mechanics based on the time reversal invariance of Schrödinger equation.

### 2.4.1. Three Accounts of Time Reversal Invariance

As we have seen in the three previous sections of this chapter, it seems that time reversal not only reverses the order of instantaneous states, but also affects the variables constituting those states, for example, it negates velocity, magnetic field and spin. So it means that the conception of time reversal presented in the introductory chapter must be modified. However, some people like Albert think no modification is needed, namely they argue that time reversal must just reverse the order of instantaneous states, leaving the variables of the states unaffected. Their main reason is that it does not make sense to change the content of the instantaneous states under time reversal, because instantaneous states are just a description of the system in a given *instant*. Logically, the descriptions of instantaneous states must be independent of the direction of time, if they are truly captured in a single instant, and not in an interval of time, as they claim [4].

According to Albert, only classical mechanics is time reversal invariant, and electromagnetism and quantum mechanics are not. In his view, the representation of the instantaneous states must just include the variables which are independent of each other. Velocity, for example, must not be included in the representation of the instantaneous states because it can be derived out of position and so is not an independent variable. Thus, in classical mechanics, instantaneous states include only position, which is invariant under time reversal, implying that classical mechanics is time reversal invariant [4].

As we have seen before, the magnetic field must be negated by negating $t$, otherwise Maxwell's equations would not be invariant under $t$-negation. But according to Albert's view, the magnetic field as an independent variable, included in the representation of instantaneous states, must remain unchanged in the reversed sequence of states. This implies that, electromagnetism is not time reversal invariant, because the Maxwell's equations are violated in the reversed sequence of states. The situation is the same in the quantum mechanics, because on the one hand, momentum and spin are required to be negated as is argued in the third step of the derivation of time reversal transformation in the previous section. But on the other hand, Albert's account of time reversal leaves these independent variables unchanged in the reverse sequence of quantum states [4].

The common objection to Albert's view is that even in a truly instantaneous state, the variables can be affected by reversing the direction of time, if they are inherently, and not necessarily explicitly, sensitive to the direction of time, as magnetic field and spin are. That is to say, in principle, a variable can to be sensitive to the direction of time even if it is not the time rate of change of another quantity. This conception of time reversal which allows one to operate on the states' variables is the most popular account among physicists and philosophers, and so is called the standard view [15].

According to this standard view, a theory is time reversal invariant if for any sequence of instantaneous states ... , $S(t_0)$, $S(t_1)$, $S(t_2)$, ... allowed by the theory, the reverse sequence of time-reversed states, ..., $S^T(t_2)$, $S^T(t_1)$, $S^T(t_0)$, ... is also allowed, where $T$ is the appropriate time reversal operator. Here, the time reversal operator $T$ is theory-dependent, and as is expected, negates velocity in classical mechanics, magnetic field in electromagnetism and momentum and spin in quantum mechanics [15].

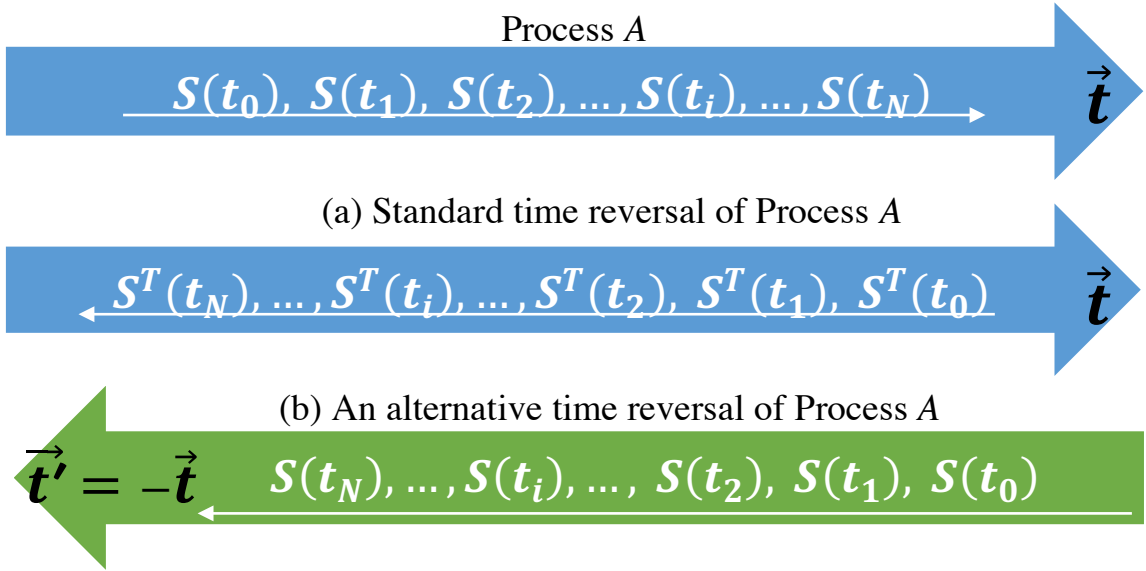

*Figure 4. Two accounts of time reversal applied on the process A: (a) Standard account, (b) Alternative account. "Time-arrows" are depicted by thick (blue or green) arrows and "process-arrows" are depicted by thin (white) arrows.*

An objection to the standard account of time reversal invariance is that $T$ somehow makes this definition trivial. It is argued that in principle, for any theory it is possible to construct a $T$ which makes the theory time reversal invariant, regardless of whether or not the theory is actually considered by the standard view as time reversal invariant [14]. In the literature, you can hardly find a real example of such $T$ suggested for a theory which is commonly regarded as time reversal non-invariant, but it seems that nothing rules out this possibility. However, this mere possibility is not considered a serious problem for the standard view. So let's explore it in more detail.

Suppose there is a finite physical process called $A$ starting at $t_0$ and ending at $t_N$. The process $A$ can be represented by subsequent instantaneous states of $S(t_i)$, where $0 < i < n$. Also suppose that we have two kinds of arrows, one of them is the process-arrow and the other one is the time-arrow. The former is depicted by a thin (white) and the latter by a thick (blue or green) arrow in figure 4. The process-arrow is just an imaginary arrow pointing to the direction in which $i$ is ascending, and the time-arrow points to the direction in which time is flowing.

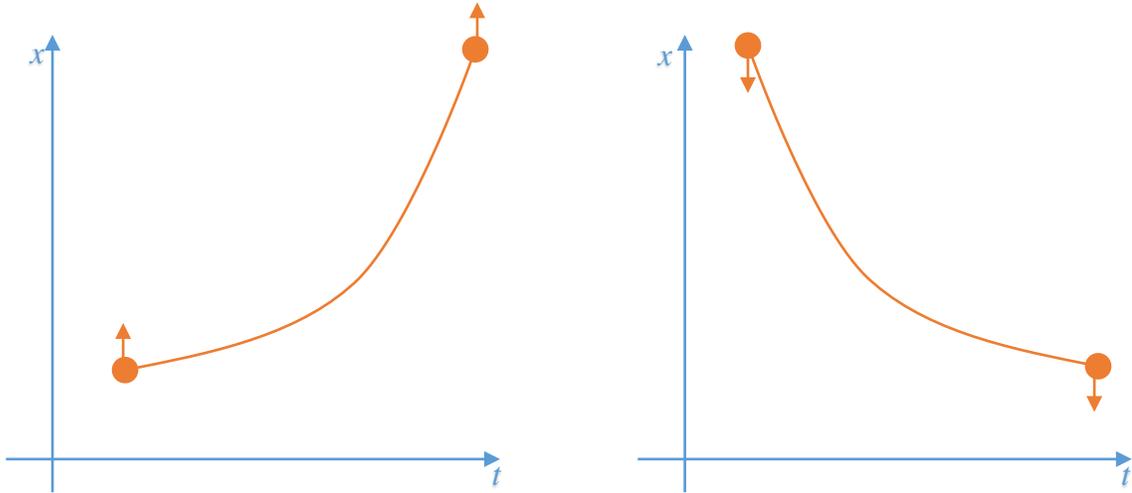

*Figure 5. Left Process: A spin-up electron moving along +**x** direction, Right Process: A spin-down electron moving along −**x** direction.*

As you see in figure 4, the top arrow in blue encompasses the subsequent instantaneous states of process $A$, such that the process-arrow and the time-arrow are parallel and both point to the right. The one below represents the time reversal of process $A$ as the standard account of time reversal suggests, i.e. the direction of time is kept fixed while the $T$ operator is applied to the inversely ordered instantaneous states of process $A$. This is why it can be said that the standard account describes a kind of reversal *in* time, not a reversal *of* time. It can be easily observed that the time-arrow and the process-arrow are antiparallel in this case. Notice that in the special condition which $T$ is the identity operator, this case would be reduced to the Albert's account of time reversal.

Figure 5 shows two quantum processes, the left one is a spin-up electron moving along +**x** direction, and the right one is a spin-down electron moving along −**x** direction. These two processes are the time reverse of each other according to the standard view, and that is why the spin of the electron is opposite in each case. However, we can have other alternative accounts of time reversal too, and I have my own, as is represented in figure 4-b. According to this alternative view, both arrows of process *A* and time must be reversed, yielding to two parallel arrows pointing to the left. That is why it can be said that this alternative account describes a kind of reversal *of* time, rather than reversal *in* time.

Thus the main difference between these two accounts of time reversal is this: according to the standard account, a process is invariant under time reversal if its reverse, interpreted in a certain sense, is also compatible with the laws of nature. But according to the alternative account, a process is invariant under time reversal if it is compatible with the laws of nature when the direction of time is reversed.

Let's define the relative direction of time as the relative direction of process-arrow to the time-arrow, and the absolute direction of time simply as the direction of the time-arrow. Given these definitions, we can say that in the standard account the relative direction of time is reversed, while the absolute direction of time is not changed. But in the alternative account, the relative direction of time is not changed, but the absolute direction of time is reversed. So, if we want to know if a certain process is sensitive to the absolute direction of time or not, the alternative view is more relevant and illuminating than the standard one.

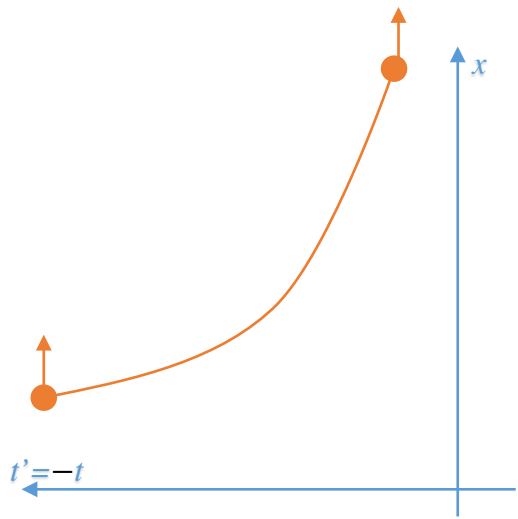

*Figure 6. Time reverse of the left process in the figure 5 according to my alternative view.*

We can apply the alternative view to the left process in figure 5. The result is depicted in figure 6, showing a spin-up electron moving along the $+\mathbf{x}$ direction. In this account, we do not have time reversal operator $T$, so the states' variables do not change under this kind of time reversal transformation. But the situation is much more complicated if we want to see what happens for the equations of different theories in the alternative account.

However, as far as the sign of $t$ is concerned, it is not difficult to determine. I think unlike the standard account, the alternative account does not require changing the sign of $t$ in the equations, because $t$ in the equations represents the relative time, and in this account the relative direction of time is not changed. But the bigger issue is that it is not guaranteed that the same relations would hold between the variables, or even that the physical constants' values would not change if the absolute direction of time were reversed. It is also possible that we may have some new variables or (internal) degrees of freedom in complex theories if the absolute direction of time is reversed.

Exploring all of these possibilities requires much work to be done. For the sake of simplicity, suppose that the constants' values and the relation between variables do not change by reversing the absolute direction of time. In the alternative account, it is not required to replace $t$ by $-t$, it can be concluded that our theories are time reversal invariant *if* they do not require any new variable to model the phenomena in a universe in which time flows in the opposite direction to ours. It is odd to imagine new variables in classical mechanics and electromagnetism because their ontology is (supposedly) not rich enough to admit them. Assuming these simplifications, it seems that classical mechanics and electromagnetism are time reversal invariant in the alternative sense.

But quantum mechanics has a rich enough ontology to admit internal degrees of freedom. For example, imagine that an electron has another spin-like internal degree of freedom, which is sensitive to the absolute direction of time and affects its behavior, such that we need to add a new variable to quantum mechanics to model it. Without assuming some weird things as the violation of the uniformity of the constants' values and the relation between variables under time reversal, we can easily imagine that quantum mechanics is time reversal *non*-invariant.

Although we cannot observe this kind of non-invariance of quantum mechanics until, for example in the above case, we can study an electron when the direction of time is actually reversed, this empirical difficulty does not mean that this account of time reversal does not deserve attention. On the contrary, I think the account of time reversal I presented here deserves serious attention, because as I tried to show, it provides us with deep insights into the rich and complex structure of quantum mechanics as our most fundamental theory.

### 2.4.2. Time Reversal Invariance of Various Interpretations of Quantum Mechanics

As we know, there are various formulations and interpretations of quantum mechanics. Although "formulation" is mostly understood as the mathematical description of a theory and "interpretation" as its ontological or philosophical description, at least for a complex theory like quantum mechanics it is too difficult to sharply distinguish them [17]. However, here I want to briefly examine time reversal invariance of the Schrödinger equation as the core equation of the wave function formulation of quantum mechanics, and I will shortly explore its implications for the three main types of the quantum mechanics' interpretations.

We saw that according to the standard account of time reversal, the natural way of thinking about time reversal invariance is this: a theory is time reversal invariant if replacing $t$ by $-t$ does not change its equation(s). For example, Newton's second law of motion is second order in time, and so will remain unchanged by that replacement, implying the time reversal invariance of the classical mechanics:

$$\underbrace{\mathbf{F} = m\mathbf{a} = \frac{d^2\mathbf{x}}{dt^2}}_{LS} \overset{t \to -t}{\Longrightarrow} \underbrace{\mathbf{F} = \frac{d^2\mathbf{x}}{(-dt)^2} = \frac{d^2\mathbf{x}}{dt^2}}_{RS}$$

$$\Rightarrow LS = RS$$

But contrary to conventional wisdom, the Schrödinger equation is not time reversal invariant in this sense, because it is first order in time:

$$\underbrace{H\psi = i\hbar\frac{\partial\psi}{\partial t}}_{LS} \overset{t \to -t}{\Longrightarrow} \underbrace{H\psi = i\hbar\frac{\partial\psi}{-\partial t} = -i\hbar\frac{\partial\psi}{\partial t}}_{RS}$$

$$\Rightarrow LS \neq RS$$

However, assuming that $H$ is *time-independent* and *real*, it can be shown that if $\psi(x, t)$ satisfies the Schrödinger equation, then $\psi^*(x, -t)$ satisfies it too:

$$\overbrace{H\psi(x, t) = i\hbar\frac{\partial\psi(x, t)}{\partial t}}^{(1)}$$

$$\Rightarrow (H\psi(x,t))^* = (i\hbar \frac{\partial \psi(x,t)}{\partial t})^*$$

$$\xrightarrow{H \ is \ real} H\psi^*(x,t) = -i\hbar \frac{\partial \psi^*(x,t)}{\partial t}$$

$$\xrightarrow{t\rightarrow -t} H\psi^*(x,-t) = -i\hbar \frac{\partial \psi^*(x,-t)}{-\partial t}$$

$$\Rightarrow \underbrace{H\psi^*(x,-t) = i\hbar \frac{\partial \psi^*(x,-t)}{\partial t}}_{(2)}$$

$$\xrightarrow{(1) \ and \ (2)} \psi(x,t) = \psi^*(x,-t)$$

The last equality entails that the Schrödinger equation would be invariant under aforementioned time reversal transformation $T = UK$, where $U$ is either $\mathbf{1}$ or $\sigma_2$ and $K$ is complex conjugation operator. In addition, it was shown before that:

$$\mathbf{Q} \xrightarrow{T} \mathbf{Q}$$

$$\mathbf{P} \xrightarrow{T} -\mathbf{P}$$

$$\boldsymbol{\sigma} \xrightarrow{T} -\boldsymbol{\sigma}$$

In favor of this sense of time reversal invariance, commonly it is argued that this transformation ($T$) is necessitated by the need to switch sign of momentum and spin under time reversal [10]. But some thinkers object to this claim by arguing that:

> There is no such necessitation. In quantum mechanics, momentum is a spatial derivative ($-i h \nabla_x$) and spin is a kind of 'space quantization'. It does not logically follow, as it does in classical mechanics, that the momentum or spin must change signs when $t \rightarrow -t$. Nor does it logically follow from $t \rightarrow -t$ that one must change $\psi \rightarrow \psi^*$ [10].

Is quantum mechanics symmetric under time reversal invariance after all? The Schrödinger equation is time reversal invariant in the modified sense mentioned above, but that does not necessarily imply quantum mechanics is time reversal invariance too. Simply, the reason is that some interpretations of the quantum mechanics modify or interrupt the Schrödinger evolution, while others do not. According to the Bell dilemma[4], there are at least three different types of interpretation of the quantum mechanics: 1) hidden variable interpretations[5], 2) collapse interpretations[6], and 3) Everett-style interpretations[7].

Let's now explore the implications of the time reversal invariance of the Schrödinger equation for the invariance of these three types of interpretations under time reversal. While the first type adds something to the Schrödinger evolution, these additions are somehow linked to the quantum state and so inherit the same invariances the quantum state does. For example, in Bohmian mechanics the particles' velocities are a function of the quantum state, and so due to the aforementioned fact that the state is not sensitive to complex conjugation, the velocities are not sensitive to the complex conjugation too.

However, the second type of interpretations are not time reversal invariant, because the Schrödinger evolution is interrupted according to these interpretations. Generally, when the wave function collapses to one of the measurement observable's eigenstates, there is no way back to the initial uncollapsed state. But interpretations of the last type will be time reversal invariant because they do not add anything to the Schrödinger evolution which is itself time reversal invariant [10].

---

[4] "Either the wave function, as given by the Schrödinger equation, is not everything, or it is not right" [17].
[5] "These are approaches that involve a denial that a quantum wave function (or any other way of representing a quantum state) yields a complete description of a physical system" [17].
[6] "These are approaches that involve modification of the dynamics to produce a collapse of the wave function in appropriate circumstances" [17].
[7] "These are approaches that reject both horns of Bell's dilemma, and hold that quantum states undergo unitary evolution at all times and that a quantum state-description is, in principle, complete" [17].

# CHAPTER III

# TIME REVERSAL INVARIANCE VIOLATION IN QUANTUM MECHANICS

According to Roberts, there are three existing ways to $T$-violation. As he states [18]:

(1) *T-Violation by Curie's Principle*. Pierre Curie declared that there is never an asymmetric effect without an asymmetric cause. This idea, together with the so-called $CPT$ theorem, provided the road to the very first detection of $T$-violation in the 20th century.

(2) *T-Violation by Kabir's Principle*. Pasha Kabir pointed that, whenever the probability of an ordinary particle decay $A \to B$ differs from that of the time-reversed decay $B' \to A'$, then we have $T$-violation. This provides a second road.

(3) *T-Violation by Wigner's Principle*. Certain kinds of matter, such as an elementary electric dipole, turn out to be $T$-violating because they have an appropriate non-degenerate energy state. This provides the final road, although it has not yet led to a successful detection of $T$-violation.

## 3.1. *T*-violation by Curie's Principle

Here, we are not interested in the original formulation of the Curie's Principle, but its version which is applicable in the quantum mechanics, as Roberts formulates it [18]:

If a quantum state fails to have a linear symmetry, then that asymmetry must also be found in either the initial state, or else in the dynamical laws.

A state has $X$ linear symmetry if it is $X$-even, i.e. it is invariant under $X$ linear transformation. The transformation $X$ is linear if $X(a\psi + b\phi) = aX\psi + aX\phi$, for any given states of $\psi$ and $\phi$. Transformations $C$, $P$, and $CP$ are linear, while as we have seen before, $T$ is anti-linear. If we have an initial state which is even/odd under one of these linear transformations, but the final state is odd/even under it, then according to the above principle, the dynamical laws governing the process violate the linear symmetry associated to it [18]. Roberts suggests a very clear mathematical formulation of this statement [18]:

> **(Unitary Curie Principle).** *Let $\mathcal{U}_t = e^{itH}$ be a continuous unitary group on a Hilbert space $\mathcal{H}$, and $R: \mathcal{H} \to \mathcal{H}$ be a linear bijection. Let $\psi_i \in \mathcal{H}$ (an "initial state") and $\psi_f = e^{itH}\psi_i$ (a "final state") for some $t \in \mathbb{R}$. If either*
>
> > *(1) (initial but not final) $R\psi_i = \psi_i$ but $R\psi_f \neq \psi_f$, or*
> >
> > *(2) (final but not initial) $R\psi_f = \psi_f$ but $R\psi_i \neq \psi_i$,*
>
> *then,*
>
> > *(3) (R-violation) $[R, H] \neq 0$.*

A real example (discovered by Cronin and Fitch in 1964): let's designate long-lived neutral $K$ mesons (kaons) by $K_L$. It is observed in the experiments that it decays to two pions ($K_L \to \pi^+\pi^-$). Also it is known that $\pi^+\pi^-$ is $CP$-odd, while $K_L$ is $CP$-even. So according to the Curie's Principle, the laws governing this decay (certain weak interactions), must be $CP$-violating [5, 18].

The next step in this way to $T$-violation is applying the $CPT$ symmetry theorem, which says that in quantum theory as is understood —describable in terms of local fields, and a unitary representation of the Poincaré group— the laws must obey $CPT$ symmetry. It means that if a certain decay of $K_L$ is $CP$-violating, it must be $T$-violating too, otherwise $CPT$ symmetry would be violated [18].

This was a very short overview of an approach to $T$-violation which employs Curie's Principle and $CPT$ theorem. Roberts mentions advantages as well as disadvantages of this approach in his view. The advantages are that it is easy and general. It is easy because one does not need know the laws (e.g. Hamiltonians or Lagrangians) connecting the initial and final states, and it is general because it can be shown that Curie's Principle is extendable to the non-unitary quantum theory, assuming that we can relax the precondition of unitarity in the application of the *CPT* theorem. The disadvantages are that it is indirect and dependent on the validity of the *CPT* theorem. It is an indirect way to *T*-violation detection because we cannot apply Curie's Principle directly to *T* as an anti-linear transformation. In principle, *CPT* symmetry can be violated in some non-standard models of particle physics making this approach inapplicable and of limited use [18].

### 3.2. *T*-violation by Kabir's Principle

Like the first way, the second way to *T*-violation is also based on a symmetry principle, called Kabir's Principle. Roberts states it [18]:

> If a transition $\psi^{in} \to \psi^{out}$ occurs with different probability than the time-reversed transition $T\psi^{out} \to T\psi^{in}$, then the laws describing those transitions must be *T*-violating.

This formulation seems precise and clear enough, so I do not go through the mathematical details. However, it should be noted that it needs unitarity as a condition. There are two successful experiments: by CPLEAR collaboration at CERN (1998) and *BABAR* collaboration at Stanford (2012) which show direct detection of *T*-violation in the *K* and *B* mesons, respectively [6, 7, 18].

The neutral *K* mesons have two special property: First, $K^0$ and $\overline{K}^0$ transform to each other repeatedly, or $K^0 \leftrightarrows \overline{K}^0$. Second, always it is possible to set the phase so that:

$$TK^0 = K^0 \text{ and } T\overline{K}^0 = \overline{K}^0 \text{ [18].}$$

This means that any kaon is time-reversed of itself (up to a phase $e^{i\varphi}$). It seems that these two properties provide us Kabir's Principle preconditions, i.e. we have both $K^0 \to \overline{K}^0$ and its reverse $\overline{K}^0 \to K^0$, while each of $K^0$ and $\overline{K}^0$ is time-reversed of itself, which means that we also have $T\overline{K}^0 \to TK^0$. According to the Kabir's Principle, if we can observe an imbalance in these two transition's occurrences, it is an actual case of $T$-violation [18, 20]. The CPLEAR experiment observing such an imbalance is described in the next part. In a general view, the story is similar in the case of $B$ mesons.

The main advantage of this method is that unlike the previous method, it provides a direct measurement of $T$-violation. Also, it does not require us to know much about the laws governing the transitions. The other advantage is that contrary to the previous method, it does not rely on the *CPT* theorem, so it could be generalized to the *CPT*-violating extensions of the standard model. But, its limitation is that we cannot apply it in cases which do not have unitary dynamics [18].

### 3.3. *T*-violation by Wigner's Principle

As Roberts says, this route to $T$-violation is not as well-known as the two previous ones. Actually, it is trying to discover some "exotic new properties of matter" [18]. He first gives a simple, but popular example in this field and then states the Wigner's Principle and discusses the physical and mathematical aspects of it.

His example is the Electric Dipole Moment (EDM): that is "the displacement between two opposite charges, or within a distribution of charges." Consider this as a property of an elementary particle. The neutron is the most well-known candidate, although this theoretically possible property has not yet been experimentally detected in it[8], nor in any other (elementary) particle [18].

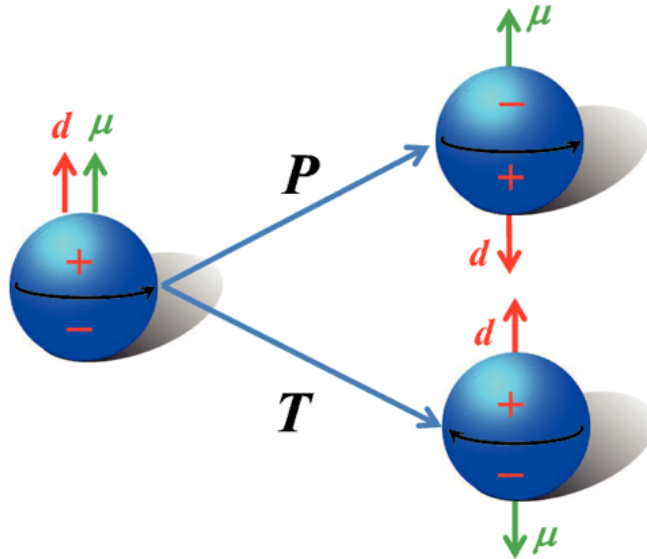

*Figure 7. A particle having EDM violates both P and T symmetries: **d** negates under P but not **μ** (top-right), and **μ** negates under T but not **d** (bottom-right). From [19].*

Consider a particle having EDM as is shown in figure 7. Such a particle both violates *T* and *P* symmetries. The simple reason is that magnetic moment (**μ**) negates under *T* but not electric moment (**d**), and electric moment negates under *P* but not magnetic moment, so in both transformations the relative direction of electric and magnetic moments to each other would be changed compared to the original situation, indicating clear cases of asymmetries. It is easy to verify that if the electric moment vanishes this argument no longer holds [18].

---

[8] The Standard Model prediction for neutron-EDM is $|\mathbf{d}| < 10^{-31} \, e. \, cm$, and the current experimental limit is $|\mathbf{d}| < 3 \times 10^{-26} \, e. \, cm$, which is negligible.

Roberts explains the $T$-violation in a slightly different way: assume $H_0$ as the interaction-free Hamiltonian of such a particle, and **J** as its angular momentum. If we put it in an electric field **E**, then the Hamiltonian describing the system will be:

$$H = H_0 + \mathbf{J} \cdot \mathbf{E}$$

By reversing time, **E** will remain unaffected, while **J** will be negated. $H$ will not be conserved under time reversal, and represents an obvious case of $T$-violation [18]. Here, the underpinning principle is Wigner's Principle [20]. Roberts states two or three somewhat different versions of it, but here I will focus on the first one because it is a little bit easier. In my opinion, his treatment is somehow confusing. Below I try to make it simpler while not missing the important details.

> **(Wigner's Principle).** If there is an eigenstate of the Hamiltonian such that: (1) that state is non-degenerate, and (2) time reversal maps that state to a different ray, then we have $T$-violation, in that $[T, H] = 0$ [18].

It is necessary to know the concepts of "degeneracy" and "ray" to understand this tricky principle [18]:

> A system is called degenerate if its Hamiltonian has distinct energy states with the same energy eigenvalue. An intuitive example of a degenerate system is the free particle on a string: the particle can either move to the left, or to the right, and have the same kinetic energy either way. When there are multiple distinct eigenstates with the same eigenvalue, those eigenstates are called degenerate states. ... But it was Wigner showed the much deeper relationship between degeneracy and time reversal invariance.
>
> Let $\mathcal{H}$ be a separable Hilbert space. A ray of $\mathcal{H}$ is a set of vectors in $\mathcal{H}$ related by a constant of unit length. We will write vectors in lower-case, and rays in upper-case Greek letters. Hence, $\Psi := \{\varphi \mid \varphi = c\psi, |c| = 1\}$ is a ray, consisting of unit multiples of the vector.

It seems that condition (1) is clear now. But condition (2) needs more clarification. The mathematical way of representing the condition (2) in a more general form, stated by Roberts, is

$$T\psi \neq e^{i\theta}\psi.$$

Mapping a state to a different ray means that $T$ non-trivially acts on the $\psi$ and multiplies it by a constant which its magnitude is not unity, and so makes the measurement probability of $T\psi$ different than measurement probability of $\psi$ [13, 18]. However, other than multiplying by such a constant, for example, $T$ can conjugate $\psi$ or affect (some of) its internal degree(s) of freedom, like reversing its spin. This is exactly what happens in the case of our time reversal operator, which conjugates the $\psi$ and negates the spins and of course, momenta [13].

Here we can rewrite the Wigner's Principle as follows [18]:

> **(Wigner's Principle).** *Let H be a self-adjoint operator on a finite-dimensional Hilbert space, which is not the zero operator. Let T be an anti-unitary bijection. If there exist an eigenvector $\psi$ of H such that,*
>
> *(1) $T\psi \neq e^{i\theta}\psi$ for any complex unit $e^{i\theta}$, and*
>
> *(2) Every eigenvector orthogonal to $\psi$ has a different eigenvalue,*
>
> *then,*
>
> *(3) (T-violation) $[T, H] \neq 0$.*

My understanding of the principle is that in a time symmetric system, $T$ cannot change the measurement probabilities. But there is an exception to this rule: in a time symmetric system having degenerate eigenstates, $T$ can change the measurement probabilities of the *degenerate* eigenstates. In this case, although the measurement probabilities might change, this does not affect the outcome of our measurements.

This is because the initial eigenstate ($\psi$) and the eigenstate which the initial eigenstate is transformed to ($T\psi$) are each other's time-flipped degenerate eigenstates having the same eigenvalue and so resulting in equal measurement outcomes. A system is time asymmetric if $T$ changes the measurement probability of a non-degenerate eigenstate of the Hamiltonian.

Thus, in a time reversal symmetric system a state and its time-flipped non-degenerate state cannot have different measurement probabilities. A time symmetric system, however, can give rise to two different measurement probabilities for the same outcome (eigenvalue) when we measure a state and its time-flipped degenerate state.

Now let's look at the Wigner's Principle proof. Roberts offers an indirect proof for it, namely he proves the contrapositive, by assuming the failure of (3), and then showing the existence of a vector violating either (1) or (2). This is the procedure: for a nonzero $h$ and some eigenvector $\psi$ of unit norm, let $H\psi = h\psi$. Recall that $T$ is anti-unitary and so has unit norm. Then,

> suppose (3) fails, and hence $[H, T] = 0$. Then $H(T\psi) = TH\psi = h(T\psi)$. This means that if $\psi$ is any eigenvector of $H$ with eigenvalue $h$, then $T\psi$ is an eigenvector with the same eigenvalue. By the spectral theorem, the eigenvectors of $H$ form an orthonormal basis set. So, since $\psi$ and $T\psi$ are both unit eigenvectors, either $T\psi = e^{i\theta}\psi$ or $\langle T\psi, \psi \rangle = 0$. The latter violates condition (2), and the former violates the condition (1). Therefore, either (1) or (2) must fail [18].

Now let's turn back to the EDM. According to Roberts, EDM is specified by these three properties [18]:

(1) (Permanence) There is an observable $D$ representing the dipole moment that is "permanent", in that $< \psi, D\psi > = a > 0$ for every eigenvector $\psi$ of the Hamiltonian $H$. Since this $\psi(\text{t})$ does not change over time except for a phase factor, permanence means that $< \psi, D\psi > = a$ has the same non-zero value for all times $t$, whence its name.

(2) (Isotropic Dynamics) Assuming that we have elementary particle, its simplest interactions are assumed to be isotropic, in that time evolution commutes with all rotations, $[e^{-itH}, R_\theta] = 0$. Note that if **J** is the "angular momentum" observable that generates the rotation $R_\theta = e^{i\theta J}$, then this is equivalent to the statement that $[T, H] = 0$.

(3) (Time Reversal Properties) Time reversal, as always, is an anti-unitary operator. It has no effect on the electric dipole observable ($TDT^{-1} = D$) when viewed as a function of position. But it does reverse the sign of angular momentum ($TJT^{-1} = -J$), since spinning things spin in the opposite orientation when their motion is reversed.

Roberts shows that whenever a particle having these properties satisfies condition (1) of the Wigner's Principle, it also satisfies the condition (2). Thus, it is resulted that if the EDM particle has non-degenerate energy eigenvectors (satisfies condition 1), it is *simply* a $T$-violating system. That is why people are very interested in the EDM particle [18].

Thus, Wigner's Principle points out a relatively easy route to $T$-violation: "If time reversal takes a non-degenerate energy eigenstate to a distinct ray, then we have $T$-violation" [18]. If a particle having EDM really exists, electromagnetic interactions suffice to violate $T$-symmetry, no need for complex behavior of weak interactions in some processes like kaon decay. But, a disadvantage is that Wigner's Principle needs us to know if a system allows a non-degenerate energy eigenstate, requiring more detailed knowledge of the Hamiltonian compared to the other two routs to $T$-violation [18].

# CHAPTER IV

# TIME REVERSAL INVARIANCE VIOLATION IN THE *K* AND *B* MESONS

In this chapter, I explain two important experiments that succeeded in the direct observation of *T*-violation in the *K* and *B* mesons. The theoretical framework of these experiments is primarily based on the Kabir's Principle introduced in the previous chapter.

## 4.1. Direct Observation of *T*-violation in the Neutral *K* Mesons

In particle physics, mixing or oscillation is a process in which a particle turns into its antiparticle and vice versa, so for the neutral *K* mesons (neutral kaons) the mixing would be $K^0 \leftrightarrows \overline{K}^0$. Mixing is governed by the weak interactions. Weak force does not conserve the strangeness quantum number (denoted by *S*), while strangeness is conserved in the strong interactions. The strangeness for $K^0$ and $\overline{K}^0$ is 1 and −1, respectively. In general, neutral kaons decay to some other final products too, so what is actually observed is an interference between mixing and decay [6, 21].

### 4.1.1. Theoretical Background

The state of the system in a given time can be shown by $|\psi> = a(t)|\text{K}^0 > + \bar{a}(t)|\overline{\text{K}}^0 >$, such that the time evolution factors $a(t)$ and $\bar{a}(t)$ satisfy the Schrödinger equation,

$$i\frac{d}{dt}\begin{pmatrix} a(t) \\ \bar{a}(t) \end{pmatrix} = \widehat{H}\begin{pmatrix} a(t) \\ \bar{a}(t) \end{pmatrix}$$

where $\widehat{H}$ is a non-Hermitian 2×2 matrix given by $\widehat{H} = M - \frac{i}{2}\Gamma$. Here *M* and $\Gamma$ are mass and decay Hermitian matrices. $K_L = \frac{1}{\sqrt{2}}(|K^0 > - |\overline{K}^0 >)$ and $K_S = \frac{1}{\sqrt{2}}(|K^0 > + |\overline{K}^0 >)$ are eigenstates of the $\hat{H}$ after diagonalization differing in lifetime ($\tau_S \approx 90$ ps and $\tau_L \approx 52$ ns $\approx 600\,\tau_S$) and mass ($\Delta m = m_L - m_S \approx 3.5 \times 10^{-12}$ MeV) [21].

The processes $K^0 \to \overline{K}^0$ and $\overline{K}^0 \to K^0$ are the time reversed of each other, so according to the Kabir's Principle, the $T$-symmetry of laws describing these processes would be violated if the occurrence probabilities of these two processes differ. To put it more precisely, considering the process $K^0 \to \overline{K}^0$ as an instantiation of $\psi^{in} \to \psi^{out}$ and its reverse $\overline{K}^0 \to K^0$ as an instantiation of $T\psi^{out} \to T\psi^{in}$, besides the aforementioned fact about kaons that $TK^0 = K^0$ and $T\overline{K}^0 = \overline{K}^0$, this is an exact and straightforward application of Kabir's Principle: conserving time symmetry needs the probability that a $K^0$ is observed as a $\overline{K}^0$ at time $\tau$ be equal to the probability that a $\overline{K}^0$ is observed as a $K^0$ at the same time $\tau$ [6, 21]. Thus, $T$-violation can be calculated according to the measure $TV$ [6, 15, 21]:

$$TV \equiv \frac{P\left(K^0 \overset{\tau}{\to} \overline{K}^0\right) - P(\overline{K}^0 \overset{\tau}{\to} K^0)}{P\left(K^0 \overset{\tau}{\to} \overline{K}^0\right) + P(\overline{K}^0 \overset{\tau}{\to} K^0)}$$

The rate of these processes is reflected in the off-diagonal elements of $\hat{H}$, so time reversal invariance needs the magnitudes of the off-diagonal matrix elements to be the same, implying,

$$\left|M_{12} - \frac{i}{2}\Gamma_{12}\right| = \left|M_{21} - \frac{i}{2}\Gamma_{21}\right| \Leftrightarrow \arg(\Gamma_{12}) - \arg(M_{12}) = n\,\pi \text{ [21]}.$$

So $T$-violation can be encoded in the small parameter $\delta_\varphi$ calculated as

$$\delta_\varphi = \pi - [\arg(\Gamma_{12}) - \arg(M_{12})].$$

We can construct another $T$-violation parameter $\varepsilon_T$ which is related to the $\delta_\varphi$ as

$$\varepsilon_T = \frac{2i\left|\hat{H}_{12}\right|^2 - \left|\hat{H}_{21}\right|^2}{2\Delta\Gamma(\Lambda_L - \Lambda_S)} = \frac{\Delta m}{2} \cdot \frac{i\Delta m + \Delta\Gamma/2}{\Delta m^2 + (\Delta\Gamma/2)^2} \cdot \delta_\varphi$$

where $\Lambda_{L,S} = \mathrm{m}_{L,S} - \frac{i}{2}\Gamma_{L,S}$ and $\Delta\Gamma = \Gamma_S - \Gamma_L$ [21].

### 4.1.2. The CPLEAR Experiment

In the CPLEAR experiment, to produce initial $K^0$ and $\overline{K}^0$, a beam of protons and antiprotons are collided:

$$p\bar{p} \rightarrow \begin{cases} K^0 K^- \pi^+ \\ \overline{K}^0 K^+ \pi^- \end{cases}$$

both reactions occur at a branching ratio of about $2 \times 10^{-3}$. Low-energy antiproton ring LEAR at CERN fired antiprotons to a gaseous hydrogen target in the center of the CPLEAR detector. The CPLEAR detector is shown in figure 8. Ten chamber layers (2 proportional chambers, 6 drift chambers, 2 streamer tubes) were used to trace charged particles resulting from annihilation and neutral-kaon decays. A 32-segment sandwich of scintillator-Cherenkov-scintillator detectors provided particle identification (kaons/ pions/ electrons). Photons were detected by an 18-layer fine-grain streamer tube/lead sampling calorimeter [21].

In total, about 100 million $K^0$ and $\overline{K}^0$ decays were reconstructed. The results refer to the analysis of the complete date set at of 70 M $K^0$, $\overline{K}^0 \rightarrow \pi^+\pi^-$ decays with $\tau > 1\tau_S$, 1.3 M $K^0$, $\overline{K}^0 \rightarrow e\pi\nu$ decays, 0.5 M $K^0$, $\overline{K}^0 \rightarrow \pi^+\pi^-\pi^0$ decays, 2 M $K^0$, $\overline{K}^0 \rightarrow \pi^0\pi^0$ decays and 17 k $K^0$, $\overline{K}^0 \rightarrow \pi^0\pi^0\pi^0$ decays [21].

In the experiment, we need to know the strangeness of the neutral kaon at its production and decay time. The initial strangeness of each neutral kaon is tagged by the charge of the charged kaon associated to it after the collision ($K^-$ is associated to $K^0$ in $K^0 K^- \pi^+$ and $K^+$ is associated to $\overline{K}^0$ in $\overline{K}^0 K^+ \pi^-$). As was mentioned before, the mere mixing of neutral kaons does not exist, because they are unstable particles and are naturally inclined to decay through some channels to more stable particles [6, 21].

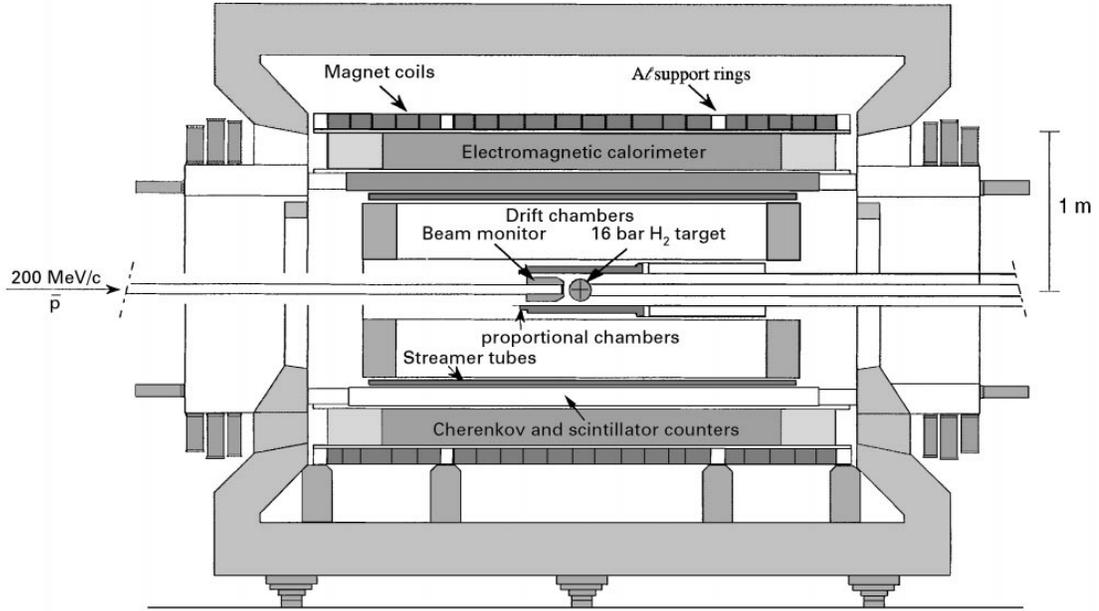

*Figure 8. View of the CPLEAR detector. From [6].*

To tag the strangeness of the kaon at the decay time $t = \tau$, we refer to semileptonic decays $K^0 \rightarrow e^+ \pi^- \nu$ and $\overline{K}^0 \rightarrow e^- \pi^+ \overline{\nu}$ which occur after kaon production. A detected positive lepton charge is associated with a $K^0$ and negative lepton charge with a $\overline{K}^0$. Here, according to Kabir's Principle, the difference in the rates ($R$) in which these processes occur is a sign of $T$-violation in the underlying laws [6, 15, 21]. Consider this ratio as the intended measurement in the experiment [6]:

$$A_T \equiv \frac{R(\overline{K}^0_{t=0} \rightarrow e^+ \pi^- \nu_{\ t=\tau_S}) - R(K^0_{t=0} \rightarrow e^- \pi^+ \overline{\nu}_{\ t=\tau_S})}{R(\overline{K}^0_{t=0} \rightarrow e^+ \pi^- \nu_{\ t=\tau_S}) + R(K^0_{t=0} \rightarrow e^- \pi^+ \overline{\nu}_{\ t=\tau_S})}$$

where $R$ denotes the rate of decay[9]. Let's represent the rate of first decay in the numerator by $\bar{N}^+$ and the second one by $N^-$. Like any other complicated experiment, there are some inefficiencies in the measurement, which we do not discuss here in detail [6, 21]. However, to compensate for these inefficiencies, two normalization factors of $< \eta > = 1.014 \pm 0.002$ and $< \xi > = 1.12023 \pm 0.00043$ are calculated for $\bar{N}^+$ and $N^-$, respectively [6, 21]:

$$A_T^{exp} = \frac{\eta \bar{N}^+ - \xi N^-}{\eta \bar{N}^+ + \xi N^-}$$

Also, there are some sources of systematic error in the measurement of $< A_T^{exp} >$, e.g. background level and background asymmetry, normalization factors, decay-time resolution, regeneration correction. The calculated amount of the systematic error is $1.0 \times 10^{-3}$ [6].

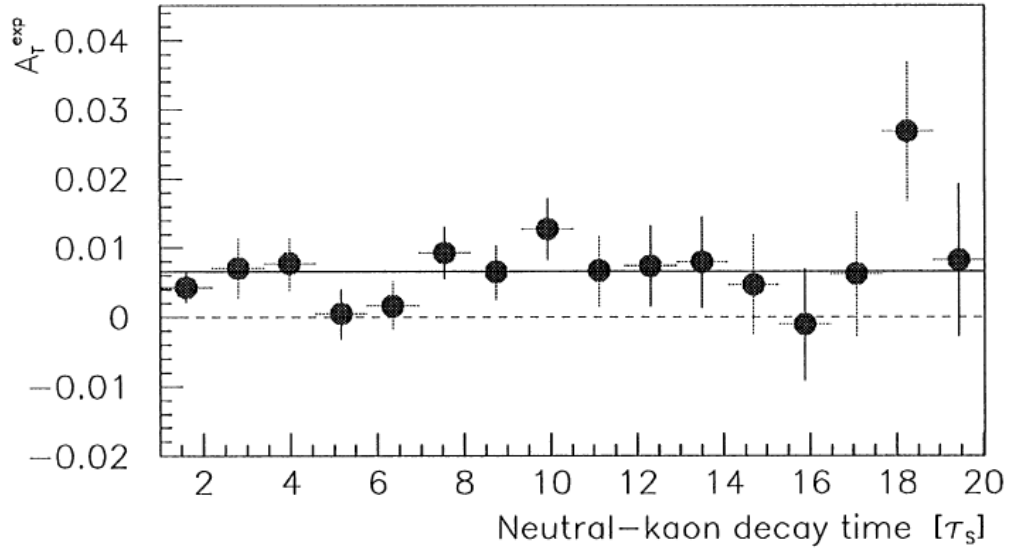

*Figure 9 . T-violation asymmetry $A_T^{exp}$ . Full line represents a constant fit in the decay-time interval 1-20$\tau_S$. From [6].*

---

[9] In the limit of $CPT$ symmetry in these decays and of the validity of the $\Delta S = \Delta Q$ rule ($S$ is the strangeness quantum number and $Q$ is the electric charge), $TV = A_T$ [6].

The measured asymmetry is shown in figure 9. The data points in the interval of $1\tau_s$ to $20\tau_s$ scatter around a positive and constant offset from zero, which represents a surplus of $\overline{K}^0 \to K^0$ process. Its average value

$$< A_T^{exp} >_{(1-20)\tau_s} = \left(6.6 \pm 1.3_{stat.} \pm 1.0_{syst.}\right) \times 10^{-3}$$

with a standard deviation $4\sigma$ represents the first direct measurement of $T$-violation [6].

## 4.2. Direct Observation of *T*-violation in the Neutral *B* Mesons

The *BABAR* collaboration using *BABAR* detector at SLAC announced the $T$-violation in the neutral $B$ mesons in 2012, namely four years after the CPLEAR experiment. The experiment involves another application of Kabir's Principle, i.e. the core idea is to compare time-dependent rates of two processes that differ by exchange of the initial and final states [7, 18].

The measurement employs the EPR effect in the entangled $B$ mesons produced in $\Upsilon(4S)$ decays. In the original experiment, the $T$-violation asymmetry in the below four independent pairs of processes which happen after $\Upsilon(4S)$ decays and satisfy Kabir's Principle is measured [7, 18].

a)  $\overline{B}^0 \leftrightarrows B_-$

b)  $B_- \leftrightarrows B^0$

c)  $\overline{B}^0 \leftrightarrows B_+$

d)  $B_+ \leftrightarrows B^0$

where $B_+$ and $B_-$ are certain orthogonal linear combinations of $B^0$ and $\overline{B}^0$. However, in the following, I just discuss one of them, namely c. Suppose that rate $R_{(\ell^+ X),(J/\psi\, K_L)}$ involves the decay of one of the neutral $B$'s into an $\ell^+ X$ state in $t_1$, and the decay of the other $B$ into $J/\psi K_L$ in $t_2$. And that rate $R_{(J/\psi K_S),(\ell^-\bar{x})}$ involves the decay of one of the neutral $B$'s into $J/\psi K_S$ in $t_1$, and the decay of the other $B$ into $\ell^-\bar{X}$ in $t_2$. Given $\Delta t = t_2 - t_1$, and under certain assumptions, this is a comparison between rates of $\bar{B}^0 \xrightarrow{\Delta t} B_+$ and $B_+ \xrightarrow{\Delta t} \bar{B}^0$ [7, 18, 22]. The asymmetry is defined as follows [22]:

$$A_T \equiv \frac{R_{(\ell^+X),(J/\psi K_L)} - R_{(J/\psi K_S),(\ell^-\bar{x})}}{R_{(\ell^+X),(J/\psi K_L)} + R_{(J/\psi K_S),(\ell^-\bar{x})}} = \frac{R\left(\bar{B}^0 \xrightarrow{\Delta t} B_+\right) - R(B_+ \xrightarrow{\Delta t} \bar{B}^0)}{R\left(\bar{B}^0 \xrightarrow{\Delta t} B_+\right) + R(B_+ \xrightarrow{\Delta t} \bar{B}^0)}$$

Figure 10 depicts the basic concept of the experiment. As you can see in the top of this figure, electron-positron collisions produce $\Upsilon(4S)$ resonances decaying to an entangled pair of $B$ mesons. When one $B$ mesons decays at $t_1$, the identity of the other is tagged while it is not measured specifically. The $B$ meson observed to decay to the final state $\ell^+X$ at $t_1$ transfers information to the other meson and dictates that it is in a $\bar{B}^0$ state. This surviving meson, tagged as $\bar{B}^0$, is observed at $t_2$ to decay into a final state $J/\psi K_L^o$ that filters the $B$ meson to be in a $B_+$ state, a linear combination of $B^0$ and $\bar{B}^0$ states. This case corresponds to a transition $\bar{B}^0 \rightarrow B_+$. To study time reversal, we have to compare the rate at which this transition occurs with the rate of the reversed transition $B_+ \rightarrow \bar{B}^0$ depicted in the bottom of the figure 10 [7, 18, 23].

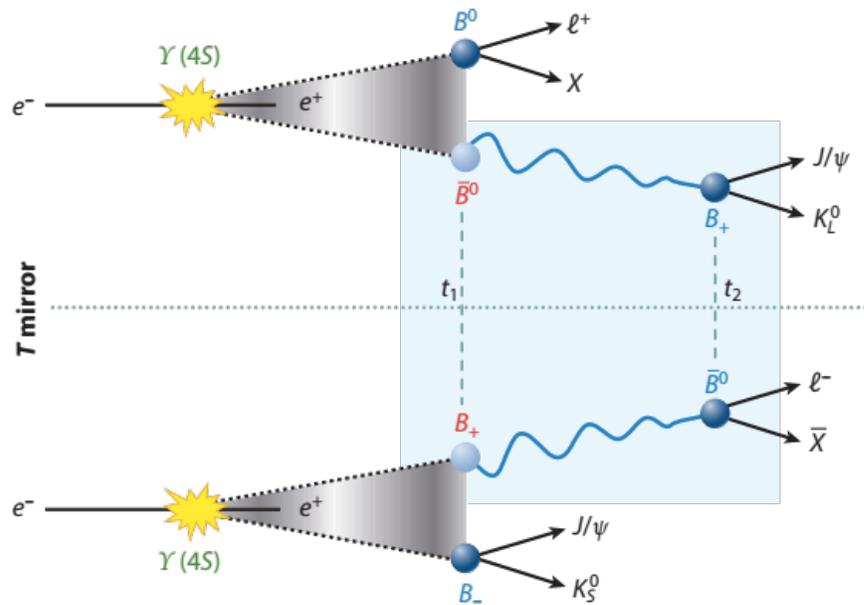

*Figure 10. Basic concept of BABAR experiment. From* [23].

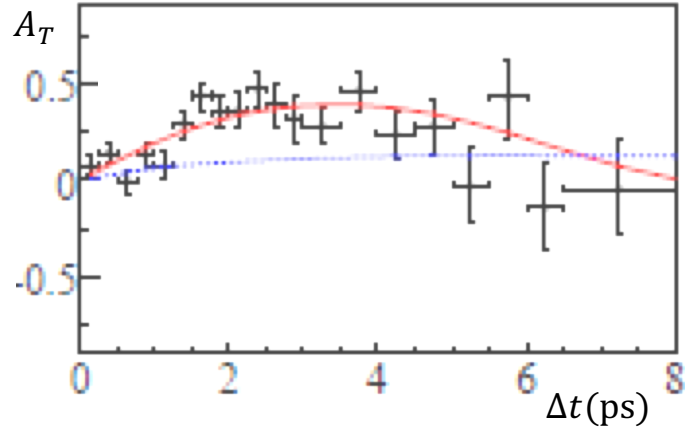

*Figure 11. The asymmetry $A_T$ measured by the BABAR experiment for transitions c. The points with error bars represent the data, the red (solid) and blue (dashed) curves represent the projections of the best fit results with and without time-reversal violation, respectively. From* [7].

$BABAR$ expresses the time dependence of the asymmetry to be of the form [7]:

$$A_T(\Delta t) \approx \frac{\Delta S_T^-}{2} sin(\Delta m_d \Delta t) + \frac{\Delta C_T^-}{2} \cos(\Delta m_d \Delta t)$$

where $\Delta m_d = 0.502 \pm 0.006 \text{ ps}^{-1}$. Using 468 million $B\bar{B}$ pairs produced in $Y(4S)$ decays collected by the $BABAR$ detector, they measured the parameters involved in $A_T$ as below [7]:

$$\Delta S_T^- = 1.17 \pm 0.18_{stat.} \pm 0.11_{syst.}$$

$$\Delta C_T^- = 0.04 \pm 0.14_{stat.} \pm 0.08_{syst.}$$

The non-zero $\Delta S_T^-$ with a significance equivalent to $14\sigma$ constitutes a direct detection of $T$-violation. Figure 11 depicts the asymmetry $A_T$ for $\Delta t \leq 8$ ps. It should be mentioned that the main difference between the CPLEAR and $BABAR$ experiments is that $K^0$ and $\bar{K}^0$ as the initial and final states of the processes under study in the CPLEAR experiment are the $CP$-conjugate of each other, while $\bar{B}^0$ and $B_+$ are not. Thus, this experiment provides the first direct observation of $T$-violation through the exchange of initial and final states in transitions that can *only* be connected by a $T$-symmetry transformation [7].

# CHAPTER V

# SUMMARY

In this thesis, I had three main goals. My first goal was giving a clear picture of spatiotemporal symmetries and their role in physics. It would not be possible to make any scientific prediction or give any scientific explanation if physical phenomena were not invariant under translation in time and space and rotation in space. As an important discovery in the history of modern physics, we have seen that contrary to initial scientists' expectations, some processes do not have spatial parity symmetry. In addition, we have seen that *CPT* symmetry (theorem), implying that physical phenomena are invariant under simultaneous transformations of charge conjugation ($C$), parity ($P$) and time reversal ($T$), has a crucial role in the Standard Model of particle physics (Chapter 1).

My second goal was introducing the standard account of time reversal invariance and studying it in classical mechanics, electromagnetism and especially quantum mechanics, as the most fundamental theory in physics. According to the standard account, a process has $T$-symmetry if its reverse, interpreted in a certain sense, is also compatible with the laws of nature. Consider two processes, a process describable solely by classical mechanics, as the elastic collision of billiard balls, and a thermodynamic process, as burning a piece of paper. If we film these processes, we can easily confirm that the first film played backward shows a process that might actually happen, but we cannot admit what the second film played backward shows might occur in the real world. Thus, according to the standard account, the first process has $T$-symmetry, but not the second.

We have seen that although as the necessary condition, it is easy to verify that Newton's second law of motion is invariant under $t \overset{T}{\to} -t$ transformation, it is not the case that all systems describable in classical mechanics have $T$-symmetry. Actually, in the Newtonian formulation of classical mechanics, the sufficient condition for $T$-symmetry is that the force must be proportional to the gradient of a *time-independent* potential. In classical mechanics, $T$ just reverses velocity $\mathbf{v}$, resulting in the motion reversal. The common belief is that electromagnetism has $T$-symmetry, too. We have seen that to guarantee this, besides current density $\mathbf{J}$ which is proportional to $\mathbf{v}$, electric field $\mathbf{E}$ must be reversed as well, otherwise Maxwell's equations would not be invariant under $t \overset{T}{\to} -t$ transformation.

As the main focus of this thesis, we studied time reversal invariance and its violation in quantum mechanics. We have seen that contrary to conventional wisdom, the Schrödinger equation is not invariant under $t \overset{T}{\to} -t$ transformation. I have shown that if $\psi(t)$ satisfies Schrödinger equation, then $\psi^*(-t)$, and not $\psi(-t)$, satisfies it as well, i.e. $T$ must conjugate the quantum state in addition to negating $t$. I briefly reviewed Robert's well-reasoned three-step plan to derive $T$ for quantum mechanics. As the outcome of his plan, we saw that $T = UK$, where $U$ is either $\mathbf{1}$ (for the position and momentum observables) or $\sigma_2$ (for the spin observable of half-spin particles), and $K$ is a complex conjugation operator. It is clear that this result is in accordance with the Schrödinger equation's invariance requirement discussed above. $T$ reverses momentum $\mathbf{P}$ and spin $\boldsymbol{\sigma}$ in quantum mechanics (Chapter 2).

Contrary to classical mechanics and electromagnetism which according to the standard account are time reversal invariant, there are some cases in which quantum mechanics violates $T$-symmetry. I reviewed three possible ways to $T$-violation in quantum mechanics. Curie's Principle provides the first way to $T$-violation: if a state having a linear symmetry is transformed to a state failing to have that symmetry, this shows that the law governing the transformation violated that symmetry. This principle together with the $CPT$ theorem were the theoretical bases for the first indirect detection of $T$-violation.

The second way is based on the Kabir's Principle, saying that if occurrences probabilities of two processes which differ by exchange of initial and final states are not equal, then the law governing these processes violates $T$. The third way to $T$-violation employs Wigner's Principle, which states that if $T$ takes a non-degenerate energy eigenstate to a distinct ray, then we have $T$-violation. This principle is applicable to an elementary particle with an electric dipole moment (EDM), however, such a particle has not yet been discovered. Contrary to the first, the last ways provide direct observations of $T$-violation, i.e. they do not depend on the $CPT$ theorem or any other intermediate principle (Chapter 3).

I explained the CPLEAR and *BABAR* experiments which based on the Kabir's Principle, directly observed $T$-violation in certain weak interactions of the neutral $K$ and $B$ mesons, respectively. In the former, to produce an initial $K^0 \overline{K}^0$, a beam of protons and antiprotons are collided. Then, the asymmetry in the rates of $K^0 \leftrightarrows \overline{K}^0$ mixing transitions is measured as a sign of $T$-violation. In the latter, electron-positron collisions produce $Y(4S)$ resonances decaying to an entangled pair of $B^0 \overline{B}^0$ mesons. The EPR effect in these mesons results in four pairs of distinct $T$-conjugated transitions. The measured asymmetry in the transitions' rates of each pair again is a direct sign of $T$-violation (Chapter 4).

My third goal was to show that there is not just a unique account of time reversal invariance in physics. Addressing this goal, I discussed two other accounts of time reversal, Albert's account and my alternative account. According to the Albert's account, $T$ must just reverse the order of instantaneous states, leaving the independent variables of the states unaffected. It was shown that under this account, classical mechanics is time reversal invariant, while electromagnetism and quantum mechanics are not. However, the standard account and Albert's account both describe a kind of reversal *of* time. Instead, my alternative account, describes a kind of reversal *in* time, that a process has $T$-symmetry if it is compatible with the laws of nature when the direction of time is reversed. I argued that even in a very simplified situation wherein $T$-symmetry of classical mechanics and electromagnetism is guaranteed, it can be violated in the quantum mechanics. I have argued that the alternative account deserves serious attention, because it provides us with deep insights into the rich and complex structure of quantum (Chapter 2).

# BIBLOGRAPHY